\documentclass[a4paper,11pt]{article}
\pdfoutput=1 

\usepackage{jcappub} 
\usepackage{array}
\usepackage[T1]{fontenc}
\usepackage{array}

\title{Parameterised Post-Newtonian Expansion in Screened Regions}

\author[a]{Ryan McManus,}
\author[a]{Lucas Lombriser,}
\author[a]{Jorge Pe\~{n}arrubia}

\affiliation[a]{Institute for Astronomy, University of Edinburgh,\\Royal Observatory, Blackford Hill, Edinburgh, EH9 3HJ, U.K.}

\emailAdd{ryanm@roe.ac.uk}
\emailAdd{llo@roe.ac.uk}
\emailAdd{jorpega@roe.ac.uk}

\newcommand{\tabspace}{\vphantom{$\frac{\frac{x^2}{x}}{\frac{x}{x}}$}}

\abstract{
The parameterised post-Newtonian (PPN) formalism has enabled stringent tests of static weak-field gravity in a theory-independent manner.
Here we incorporate screening mechanisms of modified gravity theories into the framework by introducing an effective gravitational coupling and defining the PPN parameters as functions of position.
To determine these functions we develop a general method for efficiently performing the post-Newtonian expansion in screened regimes.
For illustration, we derive all the PPN functions for a cubic galileon and a chameleon model. 
We also analyse the Shapiro time delay effect for these two models and find no deviations from General Relativity insofar as the signal path and the perturbing mass reside in a screened region of space.
}

\begin{document}
\maketitle
\flushbottom

\section{Introduction}
\label{sec:intro}

Modified gravity theories have gained interest among other reasons because of their cosmological use in explaining early universe inflation or late-time acceleration~\cite{ModifiedGravityAndCosmology,BeyondTheCosmologicalStandardModel,Cosmologicaltestsofmodifiedgravity,DarkEnergyVsModifiedGravity} (however, see refs.~\cite{NoGoTheoremsForGeneralizedChameleonFieldTheories,ChallengesToSelfAccelerationInModifiedGravityFromGravitationalWavesAndLargeScaleStructure,BreakingADarkDegeneracyWithGravitationalWaves}).
In contrast, a wide range of gravitational phenomena on scales from the Solar System to galaxies can be described by the low-energy static regime of gravity.
Because of the immediate access to this regime, low-energy tests, such as light deflection or the precession of the perihelion of Mercury, have probed relativistic gravity since the inception of General Relativity (GR).
Modern tests of gravity in the Solar System have since put very tight constraints on deviations from GR~\cite{TheConfrontationBetweenGeneralRelativityAndExperiment}.
This makes many modified gravity theories incapable of introducing significant effects on cosmological scales while simultaneously passing these stringent constraints.
However, a range of screening mechanisms have been discovered~\cite{ChameleonFieldsAwaitingSurprisesForTestsOfGravityInSpace,ScreeningLongRangeForcesthroughLocalSymmetryRestoration,ToTheProblemOfNonvanishingGravitationMass,KMOUFLAGEGravity,BeyondTheCosmologicalStandardModel} which allow for observable modifications on large scales that remain locally viable.
These mechanisms typically are due to non-linear interactions in the gravitational theory (see refs.~\cite{ClassifyingLinearlyShieldedModifiedGravityModelsInEffectiveFieldTheory,NonStandardGravitationalWavesImplyGravitationalSlip} for linear shielding effects).
These dominate in regions of high ambient matter density, such as the Solar System, and suppress deviations from GR.
However, the modifications in the low-energy static regime do not vanish completely and can still be used to constrain the range of possible modifications on cosmological scales.
The difficulty thereby lies in inferring constraints that are applicable to the wide scope of gravitational theories proposed and independent of the specifics of any theory.

In order to conduct theory independent tests of gravity in the low-energy static limit, a parameterised post-Newtonian (PPN) formalism was developed in ref.~\cite{TheoreticalFrameworksForTestingRelativisticGravityII}.
The formalism has traditionally been constructed through a linear post-Newtonian (PN) expansion of the metric field and stress-energy contributions in powers of $v/c$, where $v$ is of the order of the velocity of planets.
More specifically, the PPN formalism decomposes the metric components in terms of scalar and vector potentials which are parameterised by linear combinations of 10 constant parameters and the gravitational constant, usually set to unity.
The measured values of these parameters then capture the effects of gravity in a theory-independent way (see \cite{TheConfrontationBetweenGeneralRelativityAndExperiment} for a list of measurements).
Gravity theories can thus be mapped onto the PPN formalism by performing a low-energy static expansion of their equations of motion with their PPN parameters found as functions of theory specific parameters.
This, in principle, directly sets constraints on the theory from the measured values of the PPN parameters.
However, modified gravity theories with non-linear screening mechanisms cannot na\"ively be mapped onto the PPN formalism because the linearisation of the field equations removes contributions from the non-linear interactions which are fundamental to the screening effects.
Therefore, a simple comparison to the theory independent Solar-System tests becomes infeasible.
Moreover, screened theories do not exhibit a single low-energy limit due to the dependence of the screening effect on ambient density, giving rise to both a screened and unscreened low-energy limit.
Examples of such screening mechanisms can be split into two categories: those that depend on scalar field values controlled through potentials such as chameleon~\cite{ChameleonFieldsAwaitingSurprisesForTestsOfGravityInSpace} or symmetron screening~\cite{ScreeningLongRangeForcesthroughLocalSymmetryRestoration}; and those that depend on derivative interactions of a scalar field such as Vainshtein~\cite{ToTheProblemOfNonvanishingGravitationMass} or k-mouflage~\cite{KMOUFLAGEGravity} screening.

The PPN formulation for theories with screening has been examined previously for several cases.
Ref.~\cite{ParametrizedPostNewtonianLimitOfHorndeskisGravityTheory}, for instance, performed a low-energy static expansion for the Horndeski action with minimally coupled matter. 
This is achieved through expanding the four free functions of the Horndeski action, which depend only on the scalar field and its first derivatives, as a Taylor series.
Only the terms in the expansion linear in field perturbations are then kept.
The two PPN parameters $\gamma$, $\beta$ and the effective gravitational strength for the Horndeski action are then found in terms of Newtonian and Yukawa potentials for a spherical static system.
Due to the linearisation and hence removal of the non-linear effects, this expansion does not incorporate screening mechanisms. 
A similar expansion was performed in ref.~\cite{PostNewtonianParametersAndCosmologicalConstantOfScreenedModifiedGravity} for scalar-tensor theories in the Einstein frame with arbitrary potentials and conformal coupling functions.
Their parameterisation embeds a multitude of theories that exhibit scalar field value screening.
These mechanisms are incorporated through an effective potential that is a function of the ambient mass density. 
The minimum of this potential is used as the background value for the scalar field and is perturbed about, giving rise to an environment dependent mass.
The PPN parameters $\gamma$, $\beta$ and the effective gravitational strength are then found from a calculation that assumes a static spherical mass distribution, which again involves solutions in terms of Newtonian and Yukawa potentials.
The static mass distribution removes the effects of the vector potentials in the expansion, so that no prediction is made for several of the PPN parameters.
Furthermore, while the spherical assumption allows for an exact solution, it limits the generic nature of the expansion.  
Refs.~\cite{ClassicalDualsOfDerivativelySelfCoupledTheories} and \cite{ClassicalDualsLegendreTransformsAndTheVainshteinMechanism} used Legendre transforms and Lagrange multipliers, respectively, to find the low-energy limit in the screened regions of derivatively shielded theories.
The purpose for constructing these dual theories is to change the action into one where the expansion becomes more natural for the screened regions.
The Lagrange multiplier method was then implemented in ref.~\cite{TheParametrizedPostNewtonianVainshteinianFormalism} to find the expansion of a Vainshtein-screened cubic galileon model \cite{GalileonAsALocalModificationOfGravity} to order $(v/c)^2$.
However, the calculations are mathematically involved and are not easily generalised to more complex derivatively screened theories.
Further, this method has not been extended to include field value screening.

In order to provide a unified but systematic and efficient method for deriving the effective field equations in the limits of screening or no screening, for which one can then perform a low-energy static expansion, we have developed a scaling approach in ref.~\cite{FindingHorndeskiTheoriesWithEinsteinGravityLimits}.
The method allows one to find the dominant terms in the field equations for both screening with large field values and derivatives through the same algorithmic process.
It relies on the expansion of the scalar field around a constant background and its scaling with some exponent of the coupling constant that controls the strength of the screening term in the action.
The effective field equations describing either screening or no screening are then found by taking a formal limit of the coupling constant in the expansion of the full field equations.
This extracts the terms in the field equations that are relevant for the given limit.

In this paper, we extend our method to find a perturbative series in powers of the coupling constant around such limits.
This extended scaling method is then combined with a low-energy static expansion to produce a PPN formulation for the screened modified gravity theories.
Contrary to the more traditional formalism, the PPN parameters become functions of both time and space, as in refs.~\cite{ParametrizedPostNewtonianLimitOfHorndeskisGravityTheory,PostNewtonianParametersAndCosmologicalConstantOfScreenedModifiedGravity},
with the functional forms of the parameters found from the series of corrections in the coupling constant.
In particular, the PPN functions reproduce the GR parameters when evaluated in screened regions to leading order, while deviations are captured in higher-order corrections.
We illustrate the method by finding the expansion for a cubic galileon and chameleon model.
As an application of our results, we re-examine the implication of the Shapiro time delay measurement made by the Cassini mission~\cite{ATestOfGeneralRelativityUsingRadioLinksWithTheCassiniSpacecraft} in the context of the two screening mechanisms. 
We find that the measurement remains unaffected by the two gravitational modifications as long as the Solar System can be considered screened.
Hence, it cannot be used as a direct constraint on the models.

In section~\ref{sec:Leading Order}, we begin with a brief review of the scaling method introduced in ref.~\cite{FindingHorndeskiTheoriesWithEinsteinGravityLimits}.
We then extend the method to a higher-order perturbative series around the screened and unscreened limits in section~\ref{sec:HigherOrder}, and apply it for illustration to a cubic galileon on flat space in section~\ref{sec:cubicFlat}.
In section~\ref{sec:Cubic}, we then combine the corrections about the screened limit with a low-energy static expansion to find the metric to order $(v/c)^4$ for a curved-space cubic galileon model.
To demonstrate that our method also applies to screening by large field values, in section~\ref{sec:cham}, we also perform an expansion for a chameleon model about the screening limit to order $(v/c)^4$.
The re-examination of the Shapiro time delay for both models is presented in section~\ref{sec:Shapiro}.
Concluding remarks are made in section~\ref{sec:conclusions}.
Finally, appendix~\ref{app:PNGR} presents a primer for performing a PN expansion for GR and the PPN formalism, and the appendices \ref{app:HigherOnLower} and \ref{app:HigherLinearEquations} provide a number of technical remarks about the higher-order corrections in the scaling method.

\section{Scaling method}
\label{sec:Method}

A scaling method was developed in 
ref.~\cite{FindingHorndeskiTheoriesWithEinsteinGravityLimits} in order to find the effective equations of motion that describe a modified gravity model in the screened regime.
The method works through consistently identifying the dominant terms in the field equations when a formal limit of the coupling parameter is taken.
The key to the scaling method is that within the expansion of a field, corrections are scaled by the coupling parameter.
We define the screening limit to be a limit of the coupling parameters such that the metric field equations reproduce those of GR up to a constant effective gravitational strength.
As GR is recovered in the screened limit, higher-order corrections around this limit are required to capture the non-vanishing effects of modifying gravity.
The PPN parameters of a gravity theory in a screened region can then be found through performing a PN expansion of the effective field equations obtained at each scaling order (sections~\ref{sec:Cubic} and \ref{sec:cham}).

We briefly review the scaling method in section~\ref{sec:Leading Order}, illustrating how one can find the leading-order terms in the field equations for a given limit.
In section~\ref{sec:HigherOrder}, we then expand upon this method to find the required perturbative series around these limits.
Finally, we provide an example of this expansion with the flat-space cubic galileon in section~\ref{sec:cubicFlat}.
Note that for simplicity, in this section, we shall only consider the application of the scaling method to a scalar field equation.
For a more complete discussion that does not restrict to this flat-space limitation, we refer to ref.~\cite{FindingHorndeskiTheoriesWithEinsteinGravityLimits} (also see ref.~\cite{AParametrisationOfModifiedGravityOnNonlinearCosmologicalScales} for an application of the scaling method to further modified gravity models).

\subsection{Leading-order field equations in screened regions}
\label{sec:Leading Order}

In the following, we briefly discuss the key aspects of the scaling method developed in ref.~\citep{FindingHorndeskiTheoriesWithEinsteinGravityLimits}, which will be used extensively throughout this work. 
For illustration, consider a generic scalar field equation 
\begin{equation}
\label{eq:genericScalarEq}
\alpha^s F_u(\phi,\partial \phi)  + \alpha^t F_v(\phi,\partial \phi) = \frac{T}{M_p^2}
\end{equation}
for a scalar field $\phi$, its first derivatives, functions $F_{u,v}$, coupling constant $\alpha$, the trace of the stress-energy tensor $T$, Planck mass $M_p$ and real numbers $s$, $t$, $u$ and $v$.
It should be noted that while we restrict the discussion to first derivatives for clarity, the method trivially extends to second and higher derivatives.
We will make the separation of the scalar field 
\begin{equation}
\label{eq:PhiExpansionLeadingOrder}
\phi = \phi_0(1+\alpha^q \psi)
\end{equation}
for a constant $\phi_0$ and field perturbation $\psi$. 
Let $F_i$ scale homogeneously with respect to $\alpha^q$ with order of the subscript $i$. 
Thus, performing the scalar field expansion~\eqref{eq:PhiExpansionLeadingOrder} in the field equation~\eqref{eq:genericScalarEq}, we find
\begin{equation}
\label{eq:genericScalarEqFirstOrder}
\alpha^{s+qu} F_u(\psi,\partial\psi)  + \alpha^{t+vq} F_v(\psi,\partial\psi) = \frac{T}{M_p^2} \, .
\end{equation}
Note that at this point, the field equations are still exact.

One can now consider the family of field equations that arises from varying $\alpha$ and examine the relative importance of the different terms in the scalar field equation for the limit of large or small $\alpha$. 
The purpose of the expansion~\eqref{eq:PhiExpansionLeadingOrder} is to extract the $\alpha$ dependence from the scalar field to use it as coefficients in the field equations.
As the source is not a function of $\alpha$, it should set the scale of the dominating terms regardless of the value of $\alpha$. 
The dominating terms can be made independent of $\alpha$ though an appropriate choice of $q$ in order to balance the source upon taking the limit of large or small $\alpha$.
In so doing, we are left with a field equation where at least one other term is of the order of the source $T/M_p^2$.
This puts a restriction on the set of values that $q$ can take, denoted $Q$.
For the field equation~\eqref{eq:genericScalarEqFirstOrder}, it is clear that
\begin{equation}
q \in Q = \left\lbrace -\frac{s}{u} , -\frac{t}{v} \right\rbrace \, . 
\end{equation}

The value that $q$ takes from this set is dependent on the particular limit chosen.
In the limit of $\alpha \to \infty$, it must be that $q = \min Q \leq 0$, and for $\alpha \to 0$, $q = \max Q \geq 0$.
As we wish for the field equation~\eqref{eq:genericScalarEqFirstOrder} to be in principle solvable after taking these limits, there should be no terms that diverge. 
Thus $q$ takes the extremal values of $Q$ so that all non-vanishing exponents of $\alpha$ which appear in the field equation will be positive and negative for $\alpha\rightarrow0$ and $\alpha\rightarrow\infty$, respectively, and so the associated terms vanish when taking these limits. 

For illustration, let $-\frac{s}{u}<0<-\frac{t}{v}$, then the effective equations of motion describing the field in these limits reduce to
\begin{align}
F_v(\psi,\partial\psi) &= \frac{T}{M_p^2}\,, \ \ \ \alpha\rightarrow0\,, \\
F_u(\psi,\partial\psi) &= \frac{T}{M_p^2}\,, \ \ \ \alpha\rightarrow\infty\,,
\end{align}
after adopting the appropriate values of $q$.

The benefit of taking these limits is that it does not involve a linearisation of the equations of motion and so the non-linear features in these limits are preserved. 
As the value of $\alpha$ is fixed by theory rather than being a variable, the two formal limits should be interpreted in the context of the scalar field perturbation around $\phi_0$ such that
\begin{equation}
\label{eq:alphaApprx}
|\alpha^q \psi| \ll 1 \,
\end{equation}
and the value of $q$ being such that this combination will vanish when we take either of the limits.
When considering the tensor field of a given gravity theory, it does not need to be expanded as the scalar field in eq.~\eqref{eq:PhiExpansionLeadingOrder} but its equation of motion should contain no diverging terms for consistency.
Importantly, the tensor equation of motion will have vanishing terms in the different $\alpha$ limits. 
This is how we recover the screening mechanisms; all terms that are not present in the Einstein field equations must vanish from the modified metric field equations upon taking the appropriate $\alpha$ limit up to a constant rescaling of the gravitational constant.
There are caveats to the scaling method that are addressed in ref.~\citep{FindingHorndeskiTheoriesWithEinsteinGravityLimits} but are not relevant for this work.

\subsection{Higher-order $\alpha$-corrections}
\label{sec:HigherOrder}

The expansion \eqref{eq:PhiExpansionLeadingOrder} only describes a scalar field in the strict limits of $\alpha \to 0,\infty$.
The two field equations obtained in these limits are the leading-order descriptions of the theory in the regions where the approximation~\eqref{eq:alphaApprx} holds.
%
%
To include higher-order corrections, we perform the expansion
\begin{equation}
\label{eq:phiHigherOrder} 
\phi = \phi_0\left(1+\sum_{i=1}^\infty \alpha^{q_i} \psi^{(q_i)}\right) \, ,
\end{equation}
where $\psi^{(q_i)}$ represent the scalar field perturbations and $q_i$ are real numbers.
For this series to be perturbative, we require that the $i^{th}$ term is a smaller correction to the field than the $j^{th}$ for $i>j$, but note that this requirement alone is not sufficient for convergence.
The condition is translated into our formalism by requiring that higher-order terms in the series vanish more quickly than the lower orders when taking the limits in $\alpha$.
Hence for the limit of $\alpha\to 0$, we must insist on the ordering $q_i < q_{i+1}$ for all $i$, and similarly for $\alpha\to\infty$ we have $q_i > q_{i+1}$.
We impose that $q_1 \geq 0$ for the $\alpha \to 0$ limit and $q_1 \leq 0$ for $\alpha \to \infty$.
We denote the set of values that $q_i$ can take as $Q_i$. 

To find the equations of motion for $\psi^{(q_i)}$, we again consider a generic  homogeneous function $F_k(\phi, \partial\phi)$ with order $k$.
Importantly, the leading-order term obtained from performing the expansion~\eqref{eq:phiHigherOrder} must recover the same result as that found with the expansion~\eqref{eq:PhiExpansionLeadingOrder} in order to preserve the relevant non-linear features.
Applying the expansion~\eqref{eq:phiHigherOrder} to $F_k$ and performing a Taylor expansion in all $\alpha^{q_i}\psi^{(q_i)}$ except for $\alpha^{q_1}\psi^{(q_1)}$ gives 
\begin{align}
F_k(\phi, \partial\phi) =& F_k(\phi, \partial\phi) |_{\bar\psi =0}  \nonumber \\
&+ 
\sum_{i=2}^\infty \alpha^{q_i} \big[ \delta_{\alpha^{q_i}\psi^{(q_i)}} F_k(\phi, \partial\phi) |_{\bar\psi =0} \psi^{(q_i)} +  \delta_{\alpha^{q_i}\partial\psi^{(q_i)}} F_k(\phi, \partial\phi) |_{\bar\psi =0} \partial\psi^{(q_i)} \big] 
+ \ldots \,  \, ,
\end{align}
where the ellipses contain terms quadratic in $\psi$, $\partial\psi$ or higher and $\bar\psi = (\psi^{(q_2)},\psi^{(q_3)},\ldots)$.
At this point we recognise that $F_k(\phi, \partial\phi)|_{\bar\psi =0} = \alpha^{k q_1}F_k(\psi^{(q_1)}, \partial\psi^{(q_1)})$.
Applying this expansion to the field equation provides all terms that may contribute in the $\alpha$ limit.
For eq.~\eqref{eq:genericScalarEq}, one finds
\begin{align}
\label{eq:genericFieldEquationFullExpansion}
\frac{T}{M_p^2} = & \; \alpha^{s+u q_1} F_u(\psi^{(q_1)},\partial\psi^{(q_1)})  + \alpha^{t+vq_1} F_v(\psi^{(q_1)},\partial\psi^{(q_1)}) \nonumber \\
& + \sum_{k=u,v}\sum_{i=2}^{\infty} \alpha^{q_i} \big[ \delta_{\alpha^{q_i}\psi^{(q_i)}} F_k(\phi, \partial\phi) |_{\bar\psi =0} \psi^{(q_i)} +  \delta_{\alpha^{q_i}\partial\psi^{(q_i)}} F_k(\phi, \partial\phi) |_{\bar\psi =0} \partial\psi^{(q_i)} \big] + \ldots\,\,.
\end{align}
%
To find the equation of motion for the leading-order field perturbation $\psi^{(q_1)}$, the extremal values of $q_1$ are needed so that there exists a term independent of $\alpha$ upon taking the appropriate limit.
However, the set $Q_1$ of such values now contains values that are functions of $q_{i>1}$.
In appendix~\ref{app:HigherOnLower} we show that the extremal values of $Q_1$ are not functions of $q_{i>1}$ and that generally for all $j$ and $k$ such that $j>k$, the extremal value of $Q_j$ are only functions of $q_k$.
This implies that the values of $q_i$ can be found iteratively.
Intuitively, this follows from the ordering we have imposed on the exponents.
Thus the value for $q_1$ is found from the first line of eq.~\eqref{eq:genericFieldEquationFullExpansion}, which is equivalent to the expansion~\eqref{eq:PhiExpansionLeadingOrder}.
Hence, the leading-order solution for the full expansion coincides with the case where only one term is considered, cf.~eq.~\eqref{eq:PhiExpansionLeadingOrder}.

For illustration, let us assume that $q_1 = -s/u$ so that the equation for $\psi^{(-s/u)}$ is
\begin{equation}
 \label{eq:genericFieldEquationLeadingOrderTerm}
 F_u \big( \psi^{(-s/u)},\partial\psi^{(-s/u)} \big) = \frac{T}{M_p^2} \, .
\end{equation}
Next we aim to find the field equation for the correction $\psi^{(q_2)}$.
Using the result in appendix~\ref{app:HigherOnLower}, we see that terms that contain only functional derivatives with respect to $\psi^{(q_2)}$ or $\partial\psi^{(q_2)}$ and terms that are only functions of $\psi^{(-s/u)}$ are needed.
Inserting the solution for the leading-order term, eq.~(\ref{eq:genericFieldEquationLeadingOrderTerm}), the field equation~(\ref{eq:genericFieldEquationFullExpansion}) for the second-order correction reduces to
\begin{align}
\label{eq:genericFieldEquationFullSecondOrderExpansion}
0 = & \; \alpha^{t-vs/u} F_v(\psi,\partial\psi) \nonumber \\
& + \alpha^{q_2} \sum_{k=u,v} \big[ \delta_{\alpha^{q_2}\psi^{(q_2)}} F_k(\phi, \partial\phi) |_{\bar\psi =0} \psi^{(q_2)} +  \delta_{\alpha^{q_2}\partial\psi^{(q_2)}} F_k(\phi, \partial\phi) |_{\bar\psi =0} \partial\psi^{(q_2)} \big] + \ldots\,\,.
\end{align}
Note that there no longer exists a term that goes as $\alpha^0$, which is to be expected as such a term is associated with the leading order.
Rather, the source term for the higher order corrections must be vanishing in the strict limit as the corrections must vanish too.
Therefore, the set $Q_2$ contains the values that $q_2$ must take for a term to have the same $\alpha$ dependence as that of the slowest vanishing term.
In the second-order equation~\eqref{eq:genericFieldEquationFullSecondOrderExpansion} the slowest vanishing term goes as $\alpha^{t-vs/u}$.
We show in appendix~\ref{app:HigherLinearEquations} that only terms linear in $\psi^{(q_2)}$ or $\partial\psi^{(q_2)}$ can give the extremal values of the set $Q_2$. 
As such we may linearise the field equation \eqref{eq:genericFieldEquationFullSecondOrderExpansion} with respect to $\psi_{(q_2)}$ and its derivatives, 
\begin{align}
-\alpha^{t-vs/u} F_v(\psi,\partial\psi) = \alpha^{q_2} \sum_{k=u,v} \big[ \delta_{\alpha^{q_2}\psi^{(q_2)}} F_k(\phi, \partial\phi) |_{\bar\psi =0} \psi^{(q_2)}  +  \delta_{\alpha^{q_2}\partial\psi^{(q_2)}} F_k(\phi, \partial\phi) |_{\bar\psi =0} \partial \psi^{(q_2)}  \big] \,. 
\end{align}
In general, all higher-order corrections obey linear field equations as discussed in appendix~\ref{app:HigherLinearEquations}.
Non-linear equations are therefore restrained to the leading order, $\psi^{(q_1)}$, and how those source the higher-order corrections.

Finally, when including the tensor field, it too must be expanded in $\alpha$-corrections around the limits $\alpha\rightarrow\infty,0$. 
More specifically, we will employ the expansion 
\begin{equation}
\label{eq:gHigherOrder} 
g_{\mu\nu} = g^{(0)}_{\mu\nu} + \sum_{i=1}^{\infty} \alpha^{p_i}g^{(p_i)}_{\mu\nu}
\end{equation}
for the metric. 
We again impose that $0<p_i<p_j$ $\forall i<j$ in the limit $\alpha\to\infty$ and $0>p_i>p_j$ $\forall i<j$ for $\alpha\to 0$.
The same procedure applies for finding the metric field equations as described here for the scalar field.

\subsection{Example: flat-space cubic galileon}
\label{sec:cubicFlat}

To illustrate how to include higher-order $\alpha$-corrections in the field equations, we will examine the example of a cubic galileon model, for simplicity considered here in flat space.
We chose this example as it involves an expansion in a screened regime and also because we examine the extension to non-negligible curvature in the next section.
The field equation of the flat-space cubic galileon is given by
\begin{equation}
\label{eq:CubicFlatEom}
6\Box \phi + \alpha[(\Box \phi)^2 - (\nabla_\mu \nabla_\nu \phi)^2] = -\frac{T}{M_p^2} \, .
\end{equation} 

Applying the scaling method of ref.~\citep{FindingHorndeskiTheoriesWithEinsteinGravityLimits} at first order, one finds for the limit $\alpha \to \infty$ that $\phi = \phi_0(1+\alpha^{-\frac{1}{2}} \psi^{(-\frac{1}{2})})$, where $\psi^{(-\frac{1}{2})}$ is a solution to
\begin{equation}
\label{eq:flatCubicLeadingEOM}
\phi_0^2 [(\Box \psi^{(-\frac{1}{2})})^2 - (\nabla_\mu \nabla_\nu \psi^{(-\frac{1}{2})})^2]  = -\frac{T}{M_p^2} \, .
\end{equation}
Next, we adopt the expansion~\eqref{eq:phiHigherOrder} to find the second-order correction to this field equation, i.e., $\phi = \phi_0(1+\alpha^{-\frac{1}{2}} \psi^{(-\frac{1}{2})} + \alpha^{q_2} \psi^{(q_2)})$ with ${q_2}<-\frac{1}{2}$. 
As discussed in section \ref{sec:HigherOrder}, after performing a Taylor expansion of eq.~\eqref{eq:CubicFlatEom} in $\alpha^{q_2}\psi^{(q_2)}$, we need only examine the linear terms. 
Inserting eq.~\eqref{eq:flatCubicLeadingEOM}, the field equation for $\psi^{q_2}$ becomes
\begin{align}
\label{eq:CubicFlatEom2Alpha}
0 = & \; 3 (\alpha^{-\frac{1}{2}} \Box \psi^{(-\frac{1}{2})} + \alpha^{q_2} \Box \psi^{(q_2)}) \nonumber \\
& + \phi_0 \alpha^{\frac{1}{2}+{q_2}}[(\Box \psi^{(-\frac{1}{2})})(\Box \psi^{(q_2)}) - (\nabla_\mu \nabla_\nu \psi^{(-\frac{1}{2})})(\nabla^\mu \nabla^\nu \psi^{(q_2)})] \, .
\end{align}
We identify that the next-order source goes as $\alpha^{-\frac{1}{2}}$ and so to balance the counter-term in eq.~\eqref{eq:CubicFlatEom2Alpha}, the exponent must be $q_2 = \min \left\lbrace -\frac{1}{2}, -1 \right\rbrace  = -1$.
Using this value and taking the limit of $\alpha \to \infty$, the equation of motion for $\psi^{(-1)}$ becomes 
\begin{equation}
3 \Box \psi^{(-\frac{1}{2})} + \phi_0 [(\Box \psi^{(-\frac{1}{2})})(\Box\psi^{(-1)}) - (\nabla_\mu \nabla_\nu \psi^{(-\frac{1}{2})})(\nabla^\mu \nabla^\nu \psi^{(-1)}) ]  = 0 \, .
\end{equation}

Similarly, we can determine the second-order correction for the limit of $\alpha \to 0$. 
Briefly, to leading order, one finds $q_1=0$ and the field equation 
\begin{equation}
\label{eq:flatCubicLeadingEOMalpha0}
6\Box \psi^{(0)} = - \frac{T}{M_P^2} \, .
\end{equation}
With the expansions~\eqref{eq:phiHigherOrder} and \eqref{eq:flatCubicLeadingEOMalpha0} of the field equation, one finds that $q_2 = \max \lbrace 1, 0 \rbrace = 1$ and
the scalar field equation for $\psi^{(1)}$,
\begin{equation}
 6 \phi_0 \Box \psi^{(0)} + \phi_0^2 [(\Box \psi_1)^2 - (\nabla_\mu \nabla_\nu \psi_1)^2] = 0 \, .
\end{equation}

\section{Post-Newtonian expansion of the cubic galileon model}
\label{sec:Cubic}

To illustrate how a low-energy expansion can be performed in the screened limit of a modified gravity theory, we will first adopt the cubic galileon in the Jordan frame~\cite{GalileonAsALocalModificationOfGravity}.
The model exhibits a Vainshtein screening mechanism~\cite{ToTheProblemOfNonvanishingGravitationMass} and is the simplest of the galileon models. 
The combination of the PN expansion with our scaling method enables a systematic low-energy expansion of the model in its screened regime.
First, we employ the scaling method to obtain the screened metric field equations at both the leading and second order in $\alpha$.
Due to screening, the leading-order metric field equation is just that of GR with a constant effective gravitational coupling.
Thus, its PN expansion can be found, e.g., in ref.~\citep{TheoryAndExperimentInGravitationalPhysics} or appendix~\ref{app:PNGR}.
At second order, the screened metric field equation contains deviations from GR and so its PN expansion gives the corrections to the PPN parameters.
A benefit to adopting the cubic galileon model is that its low-energy expansion of the screened regime has previously been studied using dual Lagrangians up to PN order 2 in ref.~\citep{TheParametrizedPostNewtonianVainshteinianFormalism} and that we can directly compare our results to that source
while extending the expansion to PN order 4.

The curved space extension of the cubic galileon in the Jordan frame has the action
\begin{equation}
\label{eq:CubicAction}
S_\emph{cubic} = \frac{M_p^2}{2} \int d^4x \sqrt{-g} \left[ \phi R + \frac{2 \omega}{\phi} X - \frac{\alpha}{4}\frac{X}{\phi^3}\Box\phi \right] + S_m[g] \, ,
\end{equation}
where $S_m$ denotes the minimally coupled matter action, $\omega$ is the Brans-Dicke parameter~\cite{MachsPrincipleAndARelativisticTheoryOfGravitation}, $X\equiv-\frac{1}{2}\partial_\mu \phi \partial^\mu \phi$, $M_p^2 = (8\pi G)^{-1}$ and $\alpha$ is the coupling strength with units $mass^{-2}$.
The metric field equation is
\begin{equation}
\label{eq:CubicMetricEquation}
\phi R_{\mu \nu} = 8 \pi G \left[T_{\mu \nu} - \frac{1}{2} T g_{\mu \nu} \right] + \frac{\omega}{\phi} \nabla_\mu \phi \nabla_\nu \phi+ \frac{1}{2} \Box \phi g_{\mu \nu}+\nabla_{\mu} \nabla_\nu \phi
+ \frac{\alpha}{8} \left[ \phi^{-3} \mathcal{M}^{(3)}_{\mu\nu} + \phi^{-4} \mathcal{M}^{(4)}_{\mu\nu} \right] \, ,
\end{equation}
and the scalar field equation is 
\begin{equation}
\label{eq:CubicScalarEquation}
(3+2\omega) \Box \phi + \frac{\alpha}{4}\left[ \phi^{-2} \mathcal{S}^{(2)} + \phi^{-3} \mathcal{S}^{(3)} + \phi^{-4} \mathcal{S}^{(4)} \right] = 8 \pi G T \, .
\end{equation}
For convenience, we have made use of the rank 2 tensors~\cite{FindingHorndeskiTheoriesWithEinsteinGravityLimits}
\begin{align*}
\mathcal{M}^{(3)}_{\mu\nu} &\equiv - X \Box \phi g_{\mu \nu} - \Box \phi \nabla_\mu \phi \nabla_\nu \phi - \nabla_\mu X \nabla_\nu \phi- \nabla_\mu \phi \nabla_\nu X \, ,
\\
\mathcal{M}^{(4)}_{\mu\nu} &\equiv 6 X \nabla_\mu \phi \nabla_\nu \phi \, ,
\end{align*}
and scalar quantities~\cite{FindingHorndeskiTheoriesWithEinsteinGravityLimits} 
\begin{align*}
\mathcal{S}^{(2)} &\equiv  - (\Box \phi)^2 + (\nabla_\mu \nabla_\nu \phi)^2 - R^{\mu\nu} \nabla_\mu \phi \nabla_\nu \phi \  \,,
\\
\mathcal{S}^{(3)} &\equiv  5\nabla_\mu \phi \nabla^\mu X - X\Box \phi  \, ,
\\
\mathcal{S}^{(4)} &\equiv 18 X^2 \, .
\end{align*}

We will use the metric signature $(-,+,+,+)$.
To label both $\alpha$ and PN orders, we will write $A^{(i,j)}$ where $i$ denotes the $\alpha$ order and $j$ the PN order. 
As we will find the relevant $\alpha$ order before the PN order, we will use $A^{(i)}$ to label to the $i^{th}$ $\alpha$ order. 
This
labelling
%
should not be confused with the tensors defined above, such as $\mathcal{S}^{(2)}$, in which case the notation will read $\mathcal{S}^{(2)(i,j)}$.
This convention will hold for all composite objects such as the Ricci tensor, scalar and the d'Alembertian. 
Greek indices will run from 0 to 3, while Latin ones will run from 1 to 3.
We will denote spacial derivatives as both $\overline{\nabla}_i $ or as an index with a comma.
We let $c=1$.
Finally, we will refer to an object of PN order $(v/c)^i$ as $\mathcal{O}_\emph{PN}(i)$.

In section \ref{sec:cubicPPN2} we recover the leading-order PN expansion for the 00 and $ij$ components of the metric, and in section \ref{sec:cubicPPN34}, we find the leading-order PN expansion for the $0j$ components and the second order for the 00 component.
In section \ref{sec:CubicPPNSol}, we solve for the metric components in terms of potentials and find the PPN parameters for this theory.

\subsection{Second-order expansion for the $00$ and $ij$ components}
\label{sec:cubicPPN2}

Using the method outlined in section~\ref{sec:Leading Order}, one finds for the $\alpha\to\infty$ limit of eq.~\eqref{eq:CubicScalarEquation} that $q_1 = -\frac{1}{2}$ (see ref.~\cite{FindingHorndeskiTheoriesWithEinsteinGravityLimits} for more details).
This limit corresponds to a screened regime with field equations
\begin{align}
\label{eq:cubicScreenedMetric}
\phi_0 R^{(0)}_{\mu \nu} &= 8 \pi G \left[T^{(0)}_{\mu \nu} - \frac{1}{2} T^{(0)} g^{(0)}_{\mu \nu} \right] \, , \\
\label{eq:cubicScreenedScalar}
\frac{1}{4} \mathcal{S}^{(2)(0)} (\psi^{(-\frac{1}{2})}) &= 8 \pi G T^{(0)} \, .
\end{align}
We now wish to perform the PN expansion of these equations in $v/c$.
To do this we will expand each metric correction coming from eq.~\eqref{eq:gHigherOrder} in PN order. 
For the leading-order term we use 
\begin{equation}
g^{(0)}_{\mu\nu} \to \eta_{\mu\nu} + \sum h_{\mu\nu}^{(0, i)} \, .
\end{equation}  
As the metric field equation is just Einstein's equation with an effective gravitational constant, the field perturbations $h_{\mu\nu}^{(0, i)}$ are the same as those for the PN expansion of GR (see appendix~\ref{app:PNGR}).
This is as expected as the leading-order term should recover GR to be considered screened. 
However, the gauge conditions imposed differ due to the additional scalar field: 
\begin{align}
\label{eq:GaugeCondition1}
g^{\mu}_{i,\mu}-\frac{1}{2}g_{\mu,i}^{\mu} &= \psi_{,i} \, ,   \\
\label{eq:GaugeCondition2}
g^{\mu}_{0,\mu} - \frac{1}{2}g_{\mu,0}^{\mu} &= \psi_{,0} - \frac{1}{2}g_{00,0} \, ,
\end{align}
where indices are raised and lowered with the Minkowski metric $\eta_{\mu\nu}$. 
It is required that these conditions match terms of the same order in both $\alpha$ and PN, and so have no effect on the expansion of $g_{\mu\nu}^{(0)}$.

We are left to find the PN expansion of the leading-order scalar field equation~\eqref{eq:cubicScreenedScalar}.
Therefor, we use that the trace of the stress-energy tensor defined in eqs.~\eqref{eq:stressEnergyTensor00}-\eqref{eq:stressEnergyTensorij}, $T = -\rho$, is $\mathcal{O}_\emph{PPN}(2)$.
As it sources the scalar field equation, there must be at least one other term which is also $\mathcal{O}_\emph{PPN}(2)$. 
Using that $R_{\mu\nu}^{(0)} \approx \mathcal{O}_\emph{PN}(\geq 2)$ and $\partial_t \approx \mathcal{O}_\emph{PN}(1)$, to lowest order the scalar field equation \eqref{eq:cubicScreenedScalar} is
\begin{equation}
\label{eq:cubicPhi1}
[(\overline{\nabla}^2 \psi^{(-\frac{1}{2},i)})^2 - (\overline{\nabla}_m \overline{\nabla}_n \psi^{(-\frac{1}{2},i)})^2] = 32 \pi G \rho \, .
\end{equation}
To balance the PN orders across the equation, the field $\psi^{(-\frac{1}{2},i)}$ must be $\mathcal{O}_\emph{PN}(1)$ and so $i=1$.
Note that with equation~\eqref{eq:cubicPhi1}, we have recovered with a simple and systematic procedure the same result as presented in equation~(3.54) of ref.~\citep{TheParametrizedPostNewtonianVainshteinianFormalism}, which was obtained from consideration of a dual Lagrangian.

To determine the deviations in the metric coming from the modification of gravity, we must first find the field equations for the next order $\alpha$-correction. 
Inserting the first two terms from the expansions of the scalar field, eq.~\eqref{eq:phiHigherOrder}, and metric, eq.~\eqref{eq:gHigherOrder}, into the metric field equation~\eqref{eq:CubicMetricEquation}, we obtain 
\begin{align}
&\phi_0 \big( 1+\alpha^{-\frac{1}{2}}\psi^{(-\frac{1}{2})} \big) \big( R^{(0)}_{\mu\nu} + \alpha^{p_1} R^{(p_1)}_{\mu\nu} \big) = 8 \pi G ( \mathcal{T}^{(0)}_{\mu\nu} + \mathcal{T}^{(p_1)}_{\mu\nu} ) + \frac{\alpha^{-1}\omega\phi_0^2}{\phi_0}\partial_\mu \psi^{(-\frac{1}{2})} \partial_\nu \psi^{(-\frac{1}{2})}  \nonumber \\ 
&+\frac{1}{2}\alpha^{-\frac{1}{2}}\phi_0 \big( \Box^{(0)} + \alpha^{p_1} \Box^{(p_1)} \big) \psi^{(-\frac{1}{2})} \big( g^{(0)}_{\mu\nu} + \alpha^{p_1} g_{\mu\nu}^{(p_1)} \big) + \alpha^{-\frac{1}{2}} \phi_0  \big( \nabla^{(0)}_\mu + \alpha^{p_1} \nabla_\mu^{(p_1)} \big) \partial_\nu  \psi^{(-\frac{1}{2})} \nonumber \\
&+ \frac{\alpha}{8} \left[ \alpha^{-\frac{3}{2}} \big( M^{(3)(0)}_{\mu\nu}(\psi^{(-\frac{1}{2})}) + \alpha^{p_2} M^{(3)(p_1)}_{\mu\nu}(\psi^{(-\frac{1}{2})}) \big) \right. \nonumber\\
& \left. +  \alpha^{-2} \big( M^{(4)(0)}_{\mu\nu}(\psi^{(-\frac{1}{2})}) + \alpha^{p_1} M^{(4)(p_1)}_{\mu\nu}(\psi^{(-\frac{1}{2})}) \big) \right] \, , 
\end{align}
where $\mathcal{T}_{\mu\nu} = T_{\mu\nu} - \frac{1}{2} g_{\mu\nu} T$.
Using the lowest-order field equation, eq.~~\eqref{eq:cubicScreenedMetric}, we are left with only terms that will vanish in the limit of $\alpha \to \infty$. 
The slowest vanishing term goes as $\alpha^{-\frac{1}{2}}$ such that $p_1 = \min \lbrace -\frac{1}{2}, 0 , \frac{1}{2}\rbrace =-\frac{1}{2}$.
Taking the limit $\alpha\to \infty$ provides the second-order metric field equation 
\begin{equation}
\label{eq:secondAlphaMetric}
R^{(-\frac{1}{2})}_{\mu\nu} +  \psi^{(-\frac{1}{2})} R^{(0)}_{\mu\nu}  = 8 \pi G \phi_0^{-1} \mathcal{T}_{\mu\nu}^{(-\frac{1}{2})}+  \frac{1}{2}\Box^{(0)} \psi^{(-\frac{1}{2})} g^{(0)}_{\mu\nu} + \nabla^{(0)}_\mu  \partial_\nu  \psi^{(-\frac{1}{2})} + \frac{1}{8\phi_0} M^{(3)(0)}_{\mu\nu}[\psi^{(-\frac{1}{2})}] \, .
\end{equation}

Preforming a PN expansion of \eqref{eq:secondAlphaMetric} recovers the metric corrections arising from the modifications to gravity. 
Using that $\psi^{(-\frac{1}{2})} = \psi^{(-\frac{1}{2},1)} + \mathcal{O}_\emph{PN}(>1)$, the metric $g_{\mu\nu}^{(0)} = \eta_{\mu\nu} + h_{\mu\nu}^{(0, 2)} + \mathcal{O}_\emph{PN}(>2)$ and $g_{\mu\nu}^{(-\frac{1}{2})} =  h_{\mu\nu}^{(-\frac{1}{2}, l)} + \mathcal{O}_\emph{PN}(>l)$, the field equation to lowest order is
\begin{align}
\label{eq:secondMetricPN}
 R^{(-\frac{1}{2}, l)}_{\mu\nu}  = \frac{1}{2}\overline{\nabla}^2 \psi^{(-\frac{1}{2},1)} \eta_{\mu\nu} +  \partial_\mu  \partial_\nu  \psi^{(-\frac{1}{2},1)} \, .
\end{align}
Examining the 00 component of the metric, the equation of motion is
\begin{equation}
\label{eq:h00Half1}
-\frac{1}{2}\overline{\nabla}^2 h^{(-\frac{1}{2},l)}_{00} =  -\frac{1}{2}\overline{\nabla}^2 \psi^{(-\frac{1}{2},1)} \, . 
\end{equation}
As the right-hand side is $\mathcal{O}_\emph{PN}(1)$, so must the left-hand side and thus $l=1$. 

As in the case of GR, $h^{(-\frac{1}{2},1)}_{mn}$ should be diagonal, hence the choice of gauge in eq.~\eqref{eq:GaugeCondition1}.
Collecting the lowest-order terms, eq.~\eqref{eq:secondMetricPN} becomes
\begin{equation}
\label{eq:hmnHalf1}
-\frac{1}{2} \overline{\nabla}^2 h^{(-\frac{1}{2},1)}_{mn} =  \frac{1}{2}\overline{\nabla}^2 \psi^{(-\frac{1}{2},1)}\delta_{mn} \, .
\end{equation}
Note that eqs.~\eqref{eq:h00Half1} and \eqref{eq:hmnHalf1} are equivalent to the second part of eqs.~(3.44) and (3.49) in ref.~\citep{TheParametrizedPostNewtonianVainshteinianFormalism}, and so we recover the same results to $\mathcal{O}_\emph{PN}(2)$ as with the employment of dual Lagrangians.

\subsection{Second-order expansion for $0j$ and the third-order for $00$}
\label{sec:cubicPPN34}

To find the higher-order PN corrections to the metric, we will need the higher-order PN correction to the scalar field.
The scalar field perturbation $\psi^{(-\frac{1}{2})}$ is a solution to the field equation~\eqref{eq:cubicScreenedScalar},
\begin{equation}
\frac{1}{4} \mathcal{S}^{(2)(0)} \big[ \psi^{(-\frac{1}{2})} \big] = 8 \pi G  T^{(0)} \,,
\end{equation}
and $\psi^{(-\frac{1}{2},1)}$ satisfies the equation of motion~\eqref{eq:cubicPhi1}.
The next order PN expansion of the trace of the stress-energy tensor is of $\mathcal{O}_\emph{PN}(4)$.
Writing down all terms in eq.~\eqref{eq:cubicScreenedScalar} which are $\mathcal{O}_\emph{PN}(4)$ or linear in $\psi^{(-\frac{1}{2},k)}$, the field equation~\eqref{eq:cubicScreenedScalar} becomes 
\begin{align}
\label{eq:cubicPhi2}
& \big( \overline{\nabla}^2 \phi^{(-\frac{1}{2},1)} \big)^2 h^{i(0,2)}_i - \big( \overline{\nabla}_i\overline{\nabla}_j\phi^{(-\frac{1}{2},1)} \big)^2 h^{k(0,2)}_k -\frac{1}{4}\overline{\nabla}^2 h_{ij}^{(0,2)}\overline{\nabla}_i \psi^{(-\frac{1}{2},1)}\overline{\nabla}_j \psi^{(-\frac{1}{2},1)} \nonumber  \\
&+\big( \overline{\nabla}^2 \psi^{(-\frac{1}{2},1)} \big) \big( \overline{\nabla}^2 \psi^{(-\frac{1}{2},k)} \big) - \big( \overline{\nabla}_m \overline{\nabla}_n \psi^{(-\frac{1}{2},1)} \big) \big( \overline{\nabla}_m \overline{\nabla}_n \psi^{(-\frac{1}{2},k)} \big) = 16 \pi G \rho \left( 3 \frac{p}{\rho} - \Pi \right) \, .
\end{align}
One can see that in order to balance the PN orders in this equation, we must have that $\mathcal{O}_\emph{PN}(\psi^{(-\frac{1}{2},k)}) = \mathcal{O}_\emph{PN}(3)$.

With this relation, we are able to find the second-order PN correction to the metric for the $0j$ components and the third order of $00$.
We can then use these results to map the modified gravity theory into the PPN formalism (see appendix~\ref{app:PNGR2}).
Using the linear terms from the expansion of the $0j$ components of the Ricci tensor, eq.~\eqref{eq:R0jexpansion}, and the two gauge conditions~\eqref{eq:GaugeCondition1} and \eqref{eq:GaugeCondition2}, it is easily shown that
\begin{equation}
R_{0j} \approx -\frac{1}{2} \big( \overline{\nabla}^2 h_{0j} + \frac{1}{2}h_{00,0j} - 2 \psi_{,0j} \big)
\end{equation}
to lowest order.
The lowest PN order for the $0j$ component in equation \eqref{eq:secondAlphaMetric} is $\mathcal{O}_\emph{PN}(2)$ after inserting the expansion of the Ricci tensor, the metric and scalar field.
The resulting field equation is given by
\begin{equation}
\label{eq:h0jHalf2}
\overline{\nabla}^2 h_{0j}^{(-\frac{1}{2},2)} + \frac{1}{2}h_{00,0j}^{(-\frac{1}{2},1)}  = 0 \, . 
\end{equation}
This is equivalent to the PN expansion of the $0j$ component in GR, eq.~\eqref{eq:GRPPN0j}, but without the matter source.

Applying the gauge conditions with third-order corrections to $R^{(-\frac{1}{2})}_{00}$, eq.~\eqref{eq:Rijexpansion}, one finds that
\begin{align}
R^{(-\frac{1}{2},3)}_{00} =& - \frac{1}{2} \bigg[ \overline{\nabla}^2 h_{00}^{(-\frac{1}{2},3)} - 2 \psi_{,00}^{(-\frac{1}{2},1)} - \psi_{,j}^{(-\frac{1}{2},1)}h_{00,j}^{(0,2)} \nonumber \\
&+ 2 h_{00,j}^{(-\frac{1}{2},1)}h_{00,j}^{(0,2)} -h_{jk}^{(0,2)}h_{00,jk}^{(-\frac{1}{2},1)} -h_{jk}^{(-\frac{1}{2},1)}h_{00,jk}^{(0,2)} \bigg] \,.
\end{align}
Thus the field equation~\eqref{eq:secondAlphaMetric} to $\mathcal{O}_\emph{PN}(3)$ is
\begin{align}
\label{eq:h00half3}
 \overline{\nabla}^2 h_{00}^{(-\frac{1}{2},3)} &= \overline{\nabla}^2\psi^{(-\frac{1}{2},3)} - \psi^{(-\frac{1}{2},1)}_{,00}  - h_{00}^{(0,2)}\overline{\nabla}^2\psi^{(-\frac{1}{2},1)}   \nonumber \\
& + \psi_{,j}^{(-\frac{1}{2},1)}h_{00,j}^{(0,2)} -2h_{00,j}^{(-\frac{1}{2},1)}h_{00,j}^{(0,2)} + h_{jk}^{(-\frac{1}{2},1)}h_{00,jk}^{(0,2)} +h_{jk}^{(0,2)}h_{00,jk}^{(-\frac{1}{2},1)}  \nonumber \\ 
& - \psi^{(-\frac{1}{2},1)}\overline{\nabla}^2 h_{00}^{(0,2)}  -\frac{1}{4\phi_0}M_{00}^{(3)(0,0)} \big[ \psi^{(-\frac{1}{2},1)} \big] + 8\pi G \phi_0^{-1} \rho h_{00}^{(-\frac{1}{2},1)} \nonumber \\ 
&- \frac{1}{2} \big( 2 \partial _j h _{ij}^{(0,2)} - \partial_i h_{jj}^{(0,2)} - \partial_i h_{00}^{(0,2)} \big) \psi^{(-\frac{1}{2},1)}_{,i} \, , 
\end{align}
where the last line arises from the contribution of the Christoffel symbols.

Note that the equations for the metric components, eqs.~\eqref{eq:h00Half1}, \eqref{eq:hmnHalf1}, \eqref{eq:h0jHalf2}, and \eqref{eq:h00half3} (with exception of the term proportional to $M_{00}^{(3)(0,0)}$), are all equivalent to the corresponding field equations found for Brans-Dicke theory with no matter content~\cite{ThePostNewtonianEquationsOfHydrodynamicsInTheBransDickeTheory}.
The differences arise from different scalar field equations, eqs.~\eqref{eq:cubicPhi1} and \eqref{eq:cubicPhi2}.
That there is not matter content in eq.~\eqref{eq:h0jHalf2} results in the gauge condition~\eqref{eq:GaugeCondition2}  being trivially satisfied.

\subsection{Metric solution}
\label{sec:CubicPPNSol}

The metric components can be solved in terms of the scalar field and source terms derived from the matter content, such as the mass density or Newtonian potential.
It will prove convenient to use the notation 
\begin{align}
\big( \overline{\nabla}^2 \big)^{-1} \big[ f(\overline{x},t) \big] \equiv -\frac{1}{4\pi}\int \frac{ f(\overline{x}',t)}{|\overline{x} - \overline{x}'|} d^3x'
\end{align}
for a function $f(\overline{x},t)$.

From the equations of motion for $h_{00}^{(-\frac{1}{2},1)}$, $h_{mn}^{(-\frac{1}{2},1)}$ and $h_{0j}^{(-\frac{1}{2},2)}$, i.e., eqs.~\eqref{eq:h00Half1}, \eqref{eq:hmnHalf1} and \eqref{eq:h0jHalf2}, respectively, it follows that
\begin{align}
\label{eq:h00Half1Sol}
h_{00}^{(-\frac{1}{2},1)} & = \psi^{(-\frac{1}{2},1)} \,, \\
\label{eq:hmnHalf1Sol}
h_{mn}^{(-\frac{1}{2},1)} & = -\psi^{(-\frac{1}{2},1)}\delta_{mn} \,, \\
\label{eq:h0jHalf2Sol}
h_{0j}^{(-\frac{1}{2},2)} & = \big( \overline{\nabla}^2 \big)^{-1} \big[ - \frac{1}{2}h_{00,0j}^{(-\frac{1}{2},1)} \big]  \nonumber \\
& = \frac{1}{2} \big( \mathcal{V}^{(-\frac{1}{2},2)}_j + 3 \mathcal{W}^{(-\frac{1}{2},2)}_j +\mathcal{O}_{PN}(>2) \big) \, .
\end{align}
In the last line, we have used the identity~\eqref{eq:timeDerivativeIntergralAprox}, and in analogy to the PPN potentials $V_i$ and $W_i$ in eqs.~\eqref{eq:Vi} and \eqref{eq:Wi}, we have defined
\begin{align}
\mathcal{V}_i^{(j,k+1)} & \equiv  -\frac{1}{4\pi}\int \frac{\psi'^{(j,k)} v_j'}{|\overline{x} - \overline{x}'|^3}d^3x'     \, ,\\
\mathcal{W}_i^{(j,k+1)} & \equiv  -\frac{1}{4\pi}\int \frac{\psi'^{(j,k)}  [\overline{v}'\cdot(\overline{x}-\overline{x}')] (x-x')_j}{|\overline{x} - \overline{x}'|^5}d^3x' \, ,
\end{align}
where $\overline{v} = \partial_t \overline{x}$.

Finally, from the equation of motion for $h_{00}^{(-\frac{1}{2}, 3)}$, eq.~\eqref{eq:h00half3}, we have
\begin{align}
\label{eq:h00Half3Sol}
h_{00}^{(-\frac{1}{2},3)}  =& \ \psi^{(\frac{1}{2},3)}  
+ \tilde{\Phi}_1^{(-\frac{1}{2},3)} - 3 \mathcal{A}_\psi^{(-\frac{1}{2},3)} 
\nonumber \\
& - \mathcal{B}_\psi^{(-\frac{1}{2},3)}  + 6 G \phi_0^{-1} \tilde{\Phi}_2^{(-\frac{1}{2},3)} - \Phi_\emph{cubic}^{(-\frac{1}{2},3)} \\
 =& \ \Phi_\emph{BD}^{(-\frac{1}{2},3)} - \Phi_\emph{cubic}^{(-\frac{1}{2},3)}\,.
\end{align}
We have absorbed all terms arising from the Brans-Dicke-like part of the action into $\Phi_\emph{BD}$ and in analogy to the PPN potentials, we have defined
\begin{align}
\begin{aligned}[c]
\tilde{\Phi}_1^{(p,q+2)} &\equiv -\frac{1}{4\pi}\int \frac{\psi'^{(p,q)} v'^2}{|x-x'|^3}d^3x' \,, \\
\tilde{\Phi}_2^{(p,q+2)} &\equiv \int \frac{\rho' \psi'^{(p,q)}}{|x-x'|}d^3x' \,,  
%
\end{aligned}
\,\,\,\,
\begin{aligned}
\mathcal{A}_\psi^{(p,q+2)} &\equiv -\frac{1}{4\pi}\int \frac{\psi'^{(p,q)}  [\overline{v}'\cdot(\overline{x}-\overline{x}')]^2 }{|\overline{x} - \overline{x}'|^5}d^3x' \,, \\
\mathcal{B}_\psi^{(p,q+2)} &\equiv -\frac{1}{4\pi}\int \frac{\psi'^{(p,q)}  [\overline{a}'\cdot(\overline{x}-\overline{x}')] }{|\overline{x} - \overline{x}'|^3}d^3x' \,, 
\nonumber
\end{aligned} \\
\Phi_\emph{cubic}^{(-\frac{1}{2},3)} \equiv (\overline{\nabla}^2)^{-1} \left[ \frac{1}{4\phi_0} |\overline{\nabla} \psi^{(-\frac{1}{2},1)}|^2 \overline{\nabla}'^2 \psi^{(-\frac{1}{2},1)} \right] \,, \,\,\,\,\,\,\,\,\,\,\,\,\,\,\,\,\,\,\,\,\,\,\,\,\,\,\,\,\,\,\,\,\,\,\,\,
\end{align}
where $\overline{a} = \partial_t \overline{v}$.

In solving the metric components we have kept the scalar field explicit, rather than implementing its solution in terms of the matter source. 
This complicates our results in comparison to expansions given for other theories such as Brans-Dicke gravity \cite{TheConfrontationBetweenGeneralRelativityAndExperiment}, where the solution to the scalar field equation is used and the metric components are solved for in terms of the matter content.
However, leaving the scalar field explicit allows for a generic solution in terms of potentials derived from it, which will prove useful when examining the chameleon model in section~\ref{sec:cham}.
One may worry that the potentials $\mathcal{A}$ and $\mathcal{B}$ arise from the Brans-Dicke part of the action but have no analogue in its PN expansion. 
This is as $\psi^{(-\frac{1}{2},1)}_{,00}$ would be removed through the use of the Brans-Dicke scalar field equation.

Interestingly, the solution for $h_{00}^{(-\frac{1}{2},1)}$ in eq.~\eqref{eq:h00Half1Sol} identifies this perturbation directly with the perturbation in the scalar field of the same PN order.
Recall that the force that a test particle feels in the low-energy static limit is associated with the derivative of the 00 component of the metric.
Hence, we recognise that the inclusion of the $\alpha$-correction gives rise to a fifth force interaction between the test particle and the scalar field.

\subsection{Mapping to the parameterised post-Newtonian formalism}

Next we are left with putting all of the parts together to map the modifications to the PPN framework (see appendix~\ref{app:PNGR2}).
This constitutes changing the summation order from \emph{summing over $\alpha$ then over PN} to the \emph{summing over PN then over $\alpha$} by combining the PN summation from the GR field equations~\eqref{eq:cubicScreenedMetric} with that coming from the $\alpha$-corrections in eq.~\eqref{eq:secondAlphaMetric}. 
The results of this section will be summarised in table \ref{table:1}.

In order to match the expression to the form of the PPN metric in eq.~\eqref{eq:PPNMetric00}, the $00$ component of the metric should go as $\approx -1 +2GU$.
Hence to PN leading order, we get that 
\begin{align}
g_{00} &= -1 + 2 G \phi_0^{-1} U + \alpha^{-\frac{1}{2}}\psi^{(-\frac{1}{2},1)} \nonumber \\
&=-1 +2G_\emph{eff}U \,, 
\end{align}
where we have defined the \emph{effective gravitational coupling} 
\begin{align}
G_\emph{eff} & \equiv G^{(0)} + G^{(-\frac{1}{2})} \nonumber \\
& = G \phi_0^{-1} +  \alpha^{-\frac{1}{2}}\frac{\psi^{(-\frac{1}{2},1)}}{2U} \, .
\end{align}
Note that the correction to $G_\emph{eff}$ is controlled by the parameter $\alpha$. 

The $ij$ component of the metric must be $\approx \delta_{ij}(1+2G_\emph{eff}\gamma U)$. 
From this we can find the PPN parameter $\gamma$ as 
\begin{align}
g_{ij} &= \delta_{ij}(1+2G\phi_0^{-1}U - \alpha^{-\frac{1}{2}} \psi^{(-\frac{1}{2}),1}) \nonumber  \\
&= \delta_{ij} \left( 1+2G_\emph{eff}U \frac{ 2G^{(0)}U - \alpha^{-\frac{1}{2}}  \psi^{(-\frac{1}{2},1)}  }{ 2G^{(0)}U  + \alpha^{-\frac{1}{2}}  \psi^{(-\frac{1}{2},1)}} \right) \nonumber \\
& \equiv \delta_{ij} \left( 1+2G_\emph{eff}U \gamma \right) \, ,
\end{align}
where $\gamma$ becomes a function of space, as expected.
We have
\begin{align}
\gamma =& \gamma^{(0)} + \gamma^{(-\frac{1}{2})} \nonumber \\
=& 1 -  \frac{   2\alpha^{-\frac{1}{2}}  \psi^{(-\frac{1}{2},1)}  }{ 2G^{(0)}U  + \alpha^{-\frac{1}{2}}  \psi^{(-\frac{1}{2},1)}} \,.
\end{align}
and when taking the limit of $\alpha \to \infty$, we recover $\gamma =1$.

For the $0i$ components of the metric, the PPN expansion~\eqref{eq:PPNMetric0i} introduces several parameters: $\xi$ which is responsible for preferred location effects, $\zeta_1$ which is responsible for a breakdown in conservation of momentum, $\alpha_1$ and $\alpha_2$ which are responsible for preferred frame effects.
As this theory comes from a Lagrangian we have no reason to expect a breakdown of conservation laws, and as a scalar-tensor theory with a constant background field we expect no preferred frame effects (see ref.~\cite{SolarSystemConstraintsOnDisformalGravityTheories} for a discussion of how preferred frame effects can appear in a cosmological context).
However, we would expect preferred location effects from our expansion.
This is as the approximation~\eqref{eq:alphaApprx} is only valid for a sufficiently screened (or unscreened) region, and so the expansion favours these locations. 
This manifests as the gravitational coupling varying through space, which is responsible for preferred location effects.
Hence, we will use the parameter $\xi$ to put the metric into PPN form. 

The PPN metric for the $0i$ components is given by
\begin{equation}
\label{eq:g0iPNMetric}
g_{0i} = -\frac{1}{2}(4 \gamma +3 - 2 \xi)V_i G_\emph{eff} - \frac{1}{2}(1+ 2 \xi)W_i G_\emph{eff} \, ,
\end{equation}
where in section~\ref{sec:CubicPPNSol} we have found the metric solution 
\begin{equation}
g_{0i} = -\frac{7}{2}V_i G^{(0)} - \frac{1}{2}W_i G^{(0)} + \alpha^{-\frac{1}{2}} \big( \frac{1}{2}\mathcal{V}^{(-\frac{1}{2},3)}_i+  \frac{3}{2}\mathcal{W}^{(-\frac{1}{2},3)}_i \big) \, .
\end{equation}
For clarity , we will drop the order notation on $\mathcal{V}$ and $\mathcal{W}$.
The method used for finding the parameters $\gamma$ and $G_\emph{eff}$  in both the $00$ and $ij$ components will not work for $\xi$. 
The reason for this is that $\xi$ will need to solve two different equations coming from the prefactors of the two vector potentials.
Furthermore, $\Psi_{0i}$ may not lie in the span of $V_i$ and $W_i$ and so will have a component that cannot be absorbed into their prefactors even without this constraint.

We therefore propose here that the parameter $\xi$ should be promoted to a matrix such that $\xi_{ij} = \xi^{(0)} \delta_{ij} + \alpha^{-\frac{1}{2}} \xi^{(-\frac{1}{2})}_{ij}$. 
The reasoning behind this is that the PPN parameters are meant to indicate how much a PPN potential is transformed from one theory to another. 
For scalar potentials the most general linear transformation is scalar multiplication. However, for vector potentials we must consider a matrix acting upon the potential. 
Moreover, as we focus on screened models, the parameters of this matrix would themselves be functions of position, as for $\gamma$.
This changes eq.~\eqref{eq:g0iPNMetric} to
\begin{align}
g_{0i} &= -\frac{1}{2} \big((4 \gamma +3 )\delta_{ij} - 2 \alpha^{-\frac{1}{2}} \xi^{(-\frac{1}{2})}_{ij}  \big)V_j G_\emph{eff} - \frac{1}{2} \big(\delta_{ij} + 2\alpha^{-\frac{1}{2}} \xi^{(-\frac{1}{2})}_{ij} \big)W_j G_\emph{eff} \nonumber \\
&= -\frac{7}{2}V_i G^{(0)} - \frac{1}{2}W_i G^{(0)} + \alpha^{-\frac{1}{2}}\frac{1}{2} (\mathcal{V}_i + 3 \mathcal{W}_i) \, ,
\end{align}
where we have used that $\xi^{(0)} = 0$.
We now wish to separate this into two different matrix equations for $\xi_{ij}V_j$ and $\xi_{ij}W_j$.
There is a redundancy as there is no unique way to doing this, but the analogous form of the potentials $\mathcal{V}_i$ and $\mathcal{W}_j$ to $V_i$ and $W_j$ suggests 
\begin{align}
\label{eq:EqForChi1}
G^{(0)} \xi_{ij}^{(-\frac{1}{2})} V_j  &= \frac{1}{2} G^{(0)}  V_i \big( 4 \gamma^{(-\frac{1}{2})}  + 7 G^{(-\frac{1}{2})}/G^{(0)} \big) + \frac{1}{2}\mathcal{V}_i   \equiv G^{(0)} V_{\emph{eff}\ i} \, , \\
\label{eq:EqForChi2}
G^{(0)} \xi_{ij}^{(-\frac{1}{2})} W_j  &= -\frac{3}{2}\mathcal{W}_i - \frac{1}{2} W_i G^{(-\frac{1}{2})}   \equiv G^{(0)} W_{\emph{eff}\ i} \, ,
\end{align}
where we have defined the effective potentials $W_{\emph{eff}\ i}$ and $V_{\emph{eff}\ i}$. 
The equations~\eqref{eq:EqForChi1} and \eqref{eq:EqForChi2} are not sufficient for specifying the matrix $\xi_{ij}^{(-\frac{1}{2})}$ and another three constrains are required. 
These extra constraints need not be physical as they are a remnant of the mathematical construction we have chosen.
As such, any constraints chosen that allow for a solution will be equivalent as they only remove redundant degrees of freedom.
We impose here that the diagonal elements $\xi_{ii}^{(-\frac{1}{2})}$ vanish as this leads to an elegant solution.
We can thus solve for $\xi_{ij}^{(-\frac{1}{2})}$ to find 
\begin{equation}
\label{eq:Xi}
\xi_{ij}^{(-\frac{1}{2})} = (1-\delta_{ij}) \frac{(\epsilon_{jkl} W_k V_l)_\emph{eff}}{\epsilon_{ikl}V_k W_l} \, ,
\end{equation}
where we define $(\epsilon_{jkl} W_k V_l)_\emph{eff}$ as  $(\epsilon_{1kl} W_k V_l)_\emph{eff} = W_{eff\ 2} V_3 - V_{eff\ 2} W_3$, with analogous definitions for $(\epsilon_{2kl} W_k V_l)_\emph{eff}$ and $(\epsilon_{3kl} W_k V_l)_\emph{eff}$.
Note that this solution does not depend on the form of $V_{\emph{eff}\ i}$ or $W_{\emph{eff}\ i}$, which is specific to the theory and only requires that the denominators do not vanish.
We have only used the PPN parameter $\xi$, however we suspect that the other parameters associated with the vector potentials ($\alpha_1$, $\alpha_2$ and $\zeta_1$) would also have to be promoted to matrices to incorporate more general gravity theories.

Finally, we have the fourth-order PN corrections to the $00$ component of the metric. 
The new parameter that enters into the PPN metric at this level is $\beta$ which describes the non-linearity of gravity as it is the coupling strength of $U^2$.
The 00 component of the metric in the PPN formalism takes the form 
\begin{align}
\label{eq:PPNg00O4}
g_{00} \approx& -1 + 2G_\emph{eff}U - 2 \beta G_\emph{eff}^2 U^2  + (2\gamma +2 - 2 \xi)  G_\emph{eff} \Phi_1 \nonumber \\ 
&+ 2 (3 + \gamma - 2 \beta + 1 + \xi) G_\emph{eff}^2 \Phi_2 + 2 G_\emph{eff} \Phi_3 \nonumber \\ 
&- 2 \xi G_\emph{eff}^2 \Phi_W + 2 (\gamma - 2 \xi) G_\emph{eff} \Phi_4 + 2 \xi G_\emph{eff} \mathcal{A} \, , 
\end{align}
[or \eqref{eq:PPNMetric00}].
Since $\xi$ now appears beside a scalar, we cannot directly use the matrix $\xi_{ij}$ that we have previously defined. 
Rather, we need to use a scalar object made from $\xi_{ij}$ such that it reduces to the scalar PPN parameter $\xi$. 
The object which is most convenient to identify $\xi$ in eq.~\eqref{eq:PPNg00O4} with is $\frac{1}{3} {\rm tr} (\xi_{ij})$.
The trace will ensure that the correct order in $\alpha$ is conserved as the determinant will raise the order by a factor of four as well as ensure that the parameterisation remains linear as the determinant will multiply components of $\xi_{ij}$ together.
Since $\xi^{(0)}$ vanishes as in GR and $\xi_{ij}^{(-\frac{1}{2})}$ has vanishing trace by definition, $\frac{1}{3} {\rm tr} ( \xi_{ij})$ does not appear in eq.~\eqref{eq:PPNg00O4}.
Thus we have that 
\begin{align}
g_{00} \approx& -1 + 2G_\emph{eff}U -2 G^{(0)2} U^2 + 4 G^{(0)} \Phi_1 + 4 G^{(0)2} \Phi_2  \nonumber \\
&+ 2 G^{(0)} \Phi_3 + 6 G^{(0)} \Phi_4 + \alpha^{-\frac{1}{2}} \big( \ \Phi_\emph{BD}^{(-\frac{1}{2},3)} - \Phi_\emph{cubic}^{(-\frac{1}{2},3)} \ \big)\, .
\end{align}
From this we can extract $\beta$ such that
\begin{align}
\label{eq:Beta}
\beta &= 1 + \alpha^{-\frac{1}{2}}\left(\beta^{(-\frac{1}{2})}_\emph{BD} + \beta^{(-\frac{1}{2})}_\emph{Cubic}\right), \\
\label{eq:BDBeta}
\beta_\emph{BD}^{(-\frac{1}{2})} &= \frac{  -\Phi_\emph{BD}^{(-\frac{1}{2},3)} + G^{(-\frac{1}{2})} 
\dfrac{\delta h^{(0,4)}_{00}}{\delta G^{(0)}}
+\gamma^{(-\frac{1}{2})}
\dfrac{\delta h^{(0,4)}_{00}}{\delta \gamma^{(0)}}
}{2 G^{(0)2} U^2 + 4 G^{(0)2} \Phi_2} \, , \\
\label{eq:CubicBeta}
\beta_\emph{Cubic}^{(-\frac{1}{2})} &=  \frac{ \Phi_\emph{cubic}^{(-\frac{1}{2},3)} }{2G^{(0)2} U^2 + 4 G^{(0)2} \Phi_2}\,,
\end{align}
where we have defined $\beta_\emph{BD}$ as the terms that arise from the Brans-Dicke-like part of the action, and $\beta_\emph{cubic}$ as the term from the additional contributions of the cubic galileon model responsible for screening.
The reason for this separation will become apparent in the next section.

\section{Post-Newtonian expansion of a chameleon screened model}
\label{sec:cham}

The chameleon screening mechanism was discovered in ref.~\citep{ChameleonFieldsAwaitingSurprisesForTestsOfGravityInSpace}, where it was found that the fifth force caused by the coupling of a scalar field to matter near massive bodies can be suppressed with respect to the Newtonian force exerted by the same object.
The mechanism relies on the scalar field taking the minimum value of an effective potential, a function of the ambient mass density and the self-interaction of the field.
The mechanism is a non-linear effect, with a linearisation leading to large deviations from GR within regions that would otherwise be screened~\cite{ModelsOfFRCosmicAccelerationThatEvadeSolarSystemTests,SolarSystemConstraintsToGeneralFRGravity}.
This causes an environmentally dependent effective mass such that in regions of high density, the effective mass grows and the force becomes Yukawa suppressed.
In contrast, in regions of low density, for instance, the large-scale structure, this mass is small, and the fifth force introduces deviations from GR.  
Chameleon screening has been of particular interest in $f(R)$ gravity theories~\citep{ModelsOfFRCosmicAccelerationThatEvadeSolarSystemTests,SolarSystemConstraintsToGeneralFRGravity,FRGravityAndChameleonTheories,fRTheories}, which are equivalent to a Brans-Dicke model with $\omega=0$ and a scalar field potential~\cite{fRGravityAndScalarTensorTheory} (also see, e.g., ref.~\cite{ConstrainingChameleonModelsWithCosmology} for a review of observational constraints).
Note that the chameleon model in ref.~\citep{ChameleonFieldsAwaitingSurprisesForTestsOfGravityInSpace} is presented with an action in the Einstein frame whereas we will work entirely in the Jordan frame.
For convenience, we adopt a simple power-law potential in this frame that exhibits chameleon screening.

More specifically, we consider the Brans-Dicke-type chameleon model with the action
\begin{equation}
\label{eq:ChamAction}
S_\emph{cham}=\frac{M_p^2}{2}\int d^4x \sqrt{-g}\left[ \phi R + \frac{2 \omega}{\phi} X - \alpha(\phi-\phi_\emph{min})^n \right] + S_m[g] \,,
\end{equation}
where $0<n<1$ (see ref.~\cite{HaloModellingInChameleonTheories} for a discussion of this model).
The equations of motion for the model are given by
\begin{eqnarray}
(3+2 \omega) \Box \phi & = & M_p^{-2} T - \alpha(\phi - \phi_\emph{min})^{n-1}(2(\phi - \phi_\emph{min})-n) \,, \\
\phi R_{\mu\nu} & = & M_p^{-2} [T_{\mu\nu} - \frac{1}{2}g_{\mu\nu}T] + \frac{\omega}{\phi}\nabla_\mu \phi \nabla_\nu \phi + \frac{1}{2}\Box \phi g_{\mu\nu} + \nabla_\mu \nabla_\nu \phi + \frac{1}{2}g_{\mu\nu} \alpha (\phi - \phi_\emph{min})^n \,. \nonumber\\
\end{eqnarray}
The limit that corresponds to screening is $\alpha \to 0$ with $q = \frac{1}{1-n}$ and $\phi_0 = \phi_\emph{min}$, and the screened field equations are found to be (see ref.~\cite{FindingHorndeskiTheoriesWithEinsteinGravityLimits} for more details on the derivation)
\begin{align}
\label{eq:ChamScreenedMetric}
\phi_0 R^{(0)}_{\mu \nu} &= 8 \pi G \left[T^{(0)}_{\mu \nu} - \frac{1}{2} T^{(0)} g^{(0)}_{\mu \nu} \right] \,, \\
\label{eq:ChamSol}
\phi_0 \psi^{(\frac{1}{1-n})} &= \left(- \frac{M_p^{-2} T^{(0)} }{ n }\right)^\frac{1}{n-1} \,.
\end{align}

The PN expansion for the metric $g^{(0)}_{\mu\nu}$ at zeroth $\alpha$ order is again that of GR, as expected for a screened limit, and the leading-order $\alpha$-correction of the scalar field is
\begin{equation}
\label{eq:chamPhi1}
\phi_0 \psi^{(\frac{1}{1-n}, \frac{2}{n-1})}=\left( \frac{M_p^{-2} \rho }{ n}\right)^\frac{1}{n-1} \,.
\end{equation}
The field equation for the first metric $\alpha$-correction is given by 
\begin{align}
R^{\left(\frac{1}{1-n}\right)}_{\mu\nu} +  \psi^{\left(\frac{1}{1-n}\right)} R^{(0)}_{\mu\nu}  &= 8 \pi G \mathcal{T}^{\left(\frac{1}{1-n}\right)}_{\mu\nu}  + \frac{1}{2}\Box^{(0)} \psi^{\left(\frac{1}{1-n}\right)} g^{(0)}_{\mu\nu} \nonumber \\
&+ \nabla^{(0)}_\mu  \partial_\nu  \psi^{\left(\frac{1}{1-n}\right)} -\frac{1}{2}g_{\mu\nu}^{(0)}\phi_0^{n-1} \psi^{\left(\frac{1}{1-n}\right)\,n} \,. 
\end{align}
From a comparison to eq.~(\ref{eq:secondAlphaMetric}), one can see that the derivation of the PN expansion to $\mathcal{O}_\emph{PN}(2)$ is analogous to the calculation for the cubic galileon model.
Performing the PN expansion on the $00$, $ij$ and $0j$ components of the field equation yields 

\begin{align}
-\frac{1}{2}\nabla^2 h_{00}^{\left(\frac{1}{1-n}, \frac{2}{n-1}\right)} & = -\frac{1}{2}\nabla^2 \psi^{\left(\frac{1}{1-n}, \frac{2}{n-1}\right)} \,, \\
-\frac{1}{2}\overline{\nabla}^2 h_{mn}^{\left(\frac{1}{1-n}, \frac{2}{n-1}\right)} & = \frac{1}{2} \overline{\nabla}^2 \psi^{\left(\frac{1}{1-n}, \frac{2}{n-1}\right)}\delta_{mn} \,, \\
\label{eq:h0jHalf2Cham}
 0 & = -\frac{1}{2} (\overline{\nabla}^2 h_{0j}^{(\frac{1}{1-n},\frac{n+1}{n-1})} + \frac{1}{2}h_{00,0j}^{(\frac{1}{1-n},\frac{2}{n-1})}) \,,
\end{align}
respectively, where we have used the gauge conditions \eqref{eq:GaugeCondition1} and \eqref{eq:GaugeCondition2}.
These equations are equivalent to eqs.~\eqref{eq:h00Half1}, \eqref{eq:hmnHalf1} and \eqref{eq:h0jHalf2} but with a different leading-order scalar field solution.
The reason for recovering the same results up to the scalar field equation is that for both the chameleon and cubic galileon model, we have a Brans-Dicke-like action amended by a non-linear screening term that vanishes in the PN expansion;
the difference arises only in the scalar field profile capturing the screening mechanism.

We now consider the next-order PN correction to the scalar field equation \eqref{eq:ChamSol}. 
This is easily found to be 
\begin{equation}
\psi^{(\frac{1}{1-n}, \frac{2n}{n-1})}=\left( \frac{M_p^{-2} \rho }{ n \, \phi_0^n}\right)^\frac{1}{n-1} \frac{\Pi - 3 p / \rho} {n-1}\,.
\end{equation}
Again using the expansion for $R_{00}$ up to quadratic terms in the metric and employing the gauge conditions~\eqref{eq:GRGuage1} and \eqref{eq:GRGuage2}, we find that
\begin{align}
 \overline{\nabla}^2 h_{00}^{\left(\frac{1}{1-n}, \frac{2n}{n-1}\right)} = & \: \overline{\nabla}^2\psi^{\left(\frac{1}{1-n}, \frac{2n}{n-1}\right)} - \psi^{\left(\frac{1}{1-n}, \frac{2}{n-1}\right)}_{,00}  - h_{00}^{(0,2)} \, \overline{\nabla}^2\psi^{\left(\frac{1}{1-n}, \frac{2}{n-1}\right)}     \nonumber \\
& + \psi^{\left(\frac{1}{1-n}, \frac{2}{n-1}\right)}_{,j} \, h_{00,j}^{(0,2)} -2h_{00,j}^{\left(\frac{1}{1-n}, \frac{2}{n-1}\right)} \, h_{00,j}^{(0,2)} + h_{jk}^{\left(\frac{1}{1-n}, \frac{2}{n-1}\right)} \, h_{00,jk}^{(0,2)}   \nonumber \\ 
& +h_{jk}^{(0,2)} \, h_{00,jk}^{\left(\frac{1}{1-n}, \frac{2}{n-1}\right)} - \psi^{\left(\frac{1}{1-n}, \frac{2}{n-1}\right)} \, \overline{\nabla}^2 h_{00}^{(0,2)}  -\phi_0^{n-1} \, \psi^{\left(\frac{1}{1-n}, \frac{2}{n-1}\right)\,n} \nonumber \\
& - \frac{1}{2} \big( 2 \partial _j h _{ij}^{(0,2)} - \partial_i h_{jj}^{(0,2)} - \partial_i h_{00}^{(0,2)} \big) \, \psi^{\left(\frac{1}{1-n}, \frac{2}{n-1} \right)}_{,i} \,. 
\end{align}
As the functional form of these equations is identical to the results found for the cubic galileon, the PPN parameters $\gamma$ and $\xi$ can be defined equivalently. 
The parameter $\beta$ differs as the screening term enters into the equation through the expansion in $h_{00}$ at $\mathcal{O}_\emph{PPN}(4)$. 
As such, $\beta$ changes in eq.~\eqref{eq:Beta} by replacing $\beta_\emph{Cubic}$ with 
\begin{align}
\label{eq:ChamBeta}
\beta_\emph{Cham} &= \frac{  -\Phi_\emph{Cham}^{\left(\frac{1}{1-n}, \frac{2n}{n-1}\right)} }{2G^{(0)2} U^2 + 4 G^{(0)2} \Phi_2}\, , \\
\Phi_\emph{Cham}^{\left(\frac{1}{1-n}, \frac{2n}{n-1}\right)} &= (\overline{\nabla}^2)^{-1}[ \phi_0^{n-1} \psi^{\left(\frac{1}{1-n}, \frac{2}{n-1}\right)\,n}] \,.
\end{align}

\section{Application to Shapiro time delay and measurements of $\gamma$}
\label{sec:Shapiro}

As the presence of gravitational fields distorts space-time away from flat space, light trajectories are not straight lines in the Newtonian sense, but rather are perturbed and the time taken to travel between two points is delayed. 
The constant PPN parameter $\gamma$  was constrained by the Cassini mission to within $10^{-5}$ of the GR value ($\gamma=1$) by measuring the time delay caused by a radio echo passing the Sun~\cite{ATestOfGeneralRelativityUsingRadioLinksWithTheCassiniSpacecraft}.
This tight constraint has proven to be among the most useful tools for constraining gravitational modifications in the Solar System.

If a light ray is emitted in a weak gravitational field at a point $\overline{x} = \overline{x}_e$ at a time $t=t_e$ in a direction $\hat{n}$, the light signal will follow the path
\begin{equation}
\label{eq:pertPostion}
\overline{x}(t) = \overline{x}_e - \hat{n}(t - t_e) + \overline{x}_p(t) \,,
\end{equation}
where $\overline{x}_p(t)$ is the perturbation away from a straight line.
The Shapiro time delay of this can be found from the component of the perturbation in the direction of $\hat{n}$, $x_{p\parallel}$.
To leading order, this component of the perturbation satisfies the equation 
\begin{align}
\label{eq:dotxp}
\frac{d x_{p\parallel}}{dt} &= -\frac{1}{2}(h_{00} + h_{ii}) \\
&= - (1+ \gamma)G_\emph{eff}U \,,
\end{align}
where there is no summation over the $i$ index.

Importantly, it is this combination that is measured by the Cassini mission and
it deviates from the standard equations for these components in the traditional PPN.
This is because we allow both $G_\emph{eff}$ and $\gamma$ to be functions of position.
Recall that to leading order in $\alpha$, the value of $G^{(0)}$ and $\gamma^{(0)}$ within in the Solar System are the ones of GR and so eq.~\eqref{eq:dotxp} makes the same prediction as GR to this order. 

Using the values for $G_\emph{eff}$ and $\gamma$, see table \ref{table:1}, one finds
\begin{align}
G_\emph{eff}U(1+\gamma) =&  G \phi_0^{-1} U + \frac{1}{2} \alpha^{q} \psi^{(q,p)} +  G \phi_0^{-1} U - \frac{1}{2} \alpha^{q} \psi^{(q,p)} \nonumber \\
=& 2G\phi_0^{-1}U,
\end{align}
where $q = -\frac{1}{2}$ and $p = 1$ for the cubic galileon model and $q = \frac{1}{1-n}$ and $p = \frac{2}{n-1}$ for the chameleon model.
The combination hence contains no $\alpha$-corrections and so both the cubic galileon and chameleon models cause no deviations in the second-order $\alpha$ perturbations of the metric. 
The only deviation from the standard result is the factor of $\phi_0^{-1}$. 
However, the combination $G\phi_0^{-1}$ corresponds to the gravitational constant measured in a screened regime to leading order, eq.~\eqref{eq:cubicScreenedMetric}, and the result can thus be considered equivalent to that of the bare gravitational constant in GR.

The cancellation of modifications to the time delay also reoccurs at the next order $\alpha$ perturbation in the cubic galileon model. 
This is in agreement with the known result for Dvali-Gabadadze-Porrati (DGP) gravity~\cite{4DGravityOnABraneIn5DMinkowskiSpace} that no deviation in lensing occurs between DGP and GR~\cite{GravitationalLeakageIntoExtraDimensions}.
Hence, we expect that the cancellation also applies to all higher orders.
Similarly, it has been shown that for $f(R)$ gravity ($\omega=0$) there is no deviation in the path of light rays from GR to $\mathcal{O}_\emph{PN}(2)$~\cite{TheParamiterisedPostNewtonianLimitOfFourthOrderTheoriesOfGravity,LinearizedfRgravity} (cf.~\cite{ModelsOfFRCosmicAccelerationThatEvadeSolarSystemTests,ConstrainingfRGravityAsAScalarTensorTheory,fRTheories}), which is also consistent with our results.

Importantly, for the theories we have considered, the combination
\begin{equation}
(h_{ii} - h_{00}) = G_\emph{eff}U(1-\gamma) =  2 \alpha^q \psi^{(q,p)}
\end{equation}
is not constrained by time delay experiments.
Hence, the often quoted Cassini bound~\cite{ATestOfGeneralRelativityUsingRadioLinksWithTheCassiniSpacecraft} (applicable to standard PPN),
\begin{equation}
| \gamma - 1 |  < 2.3 \times 10 ^{-5},
 \label{eq:CassiniSTDPPN}
\end{equation}
cannot straightforwardly be employed to constrain the value of the scalar field in the Solar System for screened models.
This result further suggests that one may also want to be cautious in the interpretation of other Solar-System tests in the context of screening mechanisms. 

It should be noted, however, that local constraints can be inferred on
$f(R)$ gravity~\cite{ModelsOfFRCosmicAccelerationThatEvadeSolarSystemTests} and other chameleon models~\cite{HaloModellingInChameleonTheories} from the requirement that the
scalar field can settle from its unscreened value in the environment of the Milky Way to its
screened value in the Solar-System region or within the region of the measured Milky Way
rotation curve. The requirement of residing in a screened regime is also applicable here.

\section{Conclusions}
\label{sec:conclusions}

\begin{table}
\centering
\begin{tabular}{|c|l|c|c|}
\hline
Parameter & Interpretation & GR  & Screened MG  \\ 
\hline
$G_\emph{eff}$ & Gravitational strength. & $G$ & $G\phi_0^{-1} +  \alpha^{q}\frac{\psi^{(q,p)}}{2U}$ \tabspace\\
$\gamma$ & Amount of curvature caused by mass. & $1$ & $1 -  \frac{   2\alpha^{q}  \psi^{(q,p)}  }{ 2G\phi_0^{-1}U  + \alpha^{q}  \psi^{(q,p)}}$ \tabspace\\  
$\beta$ & Amount of non-linearity in superposition. & $1$ & $1 +\alpha^{q} \left( \beta^{(q)}_\emph{BD} + \beta^{(q)}_\emph{Scr}\right)$ \tabspace\\
$\xi_{ij}$ & Preferred location effects. & $0$ & $\alpha^{q}(1-\delta_{ij}) \frac{(\epsilon_{jkl} W_k V_l)_\emph{eff}}{\epsilon_{ikl}V_k W_l}$ \tabspace\\
\hline
\end{tabular}
%
\caption{The PPN parameters for the screened regions of the cubic galileon and chameleon models with their physical interpretations.
Contrary to their usual definitions, they are promoted to functions and the models additionally introduce a varying gravitational coupling $G_{eff}$.
The scalar field perturbations $\psi$ are solutions to the leading-order scalar field equations~\eqref{eq:cubicPhi1} for the cubic galileon model, where $q=-\frac{1}{2}$, $p = 1$, and \eqref{eq:chamPhi1} for the chameleon model, where $q=\frac{1}{1-n}$, $p=\frac{2}{n-1}$.
Furthermore, $\alpha$ is the coupling to the screening terms in the action, used as the scaling parameter, $V_i$ and $W_j$ are the usual PPN vector potentials, and $(\epsilon_{jkl} W_k V_l)_\emph{eff}$ is defined in eq.~\eqref{eq:Xi}.
Finally, $\beta_\emph{BD}$  is a complicated function given by eq.~\eqref{eq:BDBeta} which is shared between the cubic galileon and chameleon models and $\beta_\emph{Scr}$ is either $\beta_\emph{Cubic}$ or $\beta_\emph{Cham}$, which are specified for each theory in eqs.~\eqref{eq:CubicBeta} and \eqref{eq:ChamBeta}.
}
\label{table:1}
\end{table}

We have demonstrated how an efficient and systematic PPN expansion can be performed in the screened regimes of different modified gravity models with fundamentally different screening mechanisms.
For this purpose, we extend the scaling method developed in ref.~\cite{FindingHorndeskiTheoriesWithEinsteinGravityLimits} to a higher-order expansion that can be applied in the screened or unscreened limits of the field equations of a given theory.
This can then be combined with a low-energy, static expansion of the metric and scalar fields in these regimes.
In particular, the procedure finds the PPN parameters for screened modified gravity models along with an effective gravitational coupling, which are generalised to time and space dependent functions. 
Moreover, we propose that the PPN parameter $\xi$, characterising preferred location effects, should be promoted to a matrix that rotates and stretches the PPN vector potentials.
This is so that the expansion of the screened models can be placed into to the PPN framework without introducing new, theory-dependent potentials.
Our method results in the PPN parameters being expressed as a series where the size of successive terms is controlled by the coupling constant associated with the screening term in the action of the modified gravity theory.

We apply our method to calculate the PPN parameters for a cubic galileon and chameleon model and summarise our results in table~\ref{table:1}.
As expected due to screening, we find that these theories recover GR at leading order.
We then compute the first-order correction to these parameters as functions of the scalar field.
The functional form for these corrections to the PPN parameters are equivalent among the two gravity models, which is attributed to the Brans-Dicke-type action adopted in both cases.
The corrections themselves, however, differ due to the scalar field solving different equations of motion.
As a further application of our method, we re-examine measurements of the Shapiro time delay.
We find that the constraint traditionally quoted for the PPN parameter $\gamma$ cannot directly be applied to modifications of gravity with screening effects.
In particular, this is attributed to the generalised effective gravitational strength and $\gamma$ combining to cancel deviations in observables from the predictions of GR.
We explicitly show this for the cubic galileon and chameleon model, finding the time delay in screened regions predicted by both theories to be identical to that of GR. 
As a result, bounds on $\gamma$ found from time delay experiments in the Solar System such as the Cassini mission~\cite{ATestOfGeneralRelativityUsingRadioLinksWithTheCassiniSpacecraft} do not constrain these theories directly.
That is, there are no deviations from GR as long as the Solar System can be considered screened.
Hence, future tests of $\gamma$ that depend on the time delay of light (see, e.g., ref.~\cite{SolarSystemTestsOfTheRelativisticGravity} for a review) will not trivially constrain these theories, regardless of accuracy, and more subtle considerations need to be made.

The natural extension to this work will be the application of our method to the Horndeski action~\cite{SecondOrderScalarTensorFieldEquationsInAFourDimensionalSpace}, in particular the subspace of models that allow for a screened Einstein limit~\cite{FindingHorndeskiTheoriesWithEinsteinGravityLimits}.
This seems feasible since in our method the equations of motion for the higher-order corrections become linear with the non-linearities restricted to the leading order and the background terms of each correction.
Thus, the corrections to the metric and scalar field remain functions of known parameters and the leading-order solutions, which also contain the field profile required for screening.
A caveat of our approach so far, however, is that in finding the PPN parameters, we have restricted to time independent background fields, which excludes screening effects relying on an evolving cosmological background (see, e.g., ref.~\cite{DisformalTheoriesOfGravity}). 
One may also wish to examine the mathematical nature of the expansion in more detail to investigate its converge as well as the assumed commutativity of the $\alpha$ expansion and the PN expansion.
Finally, the expansion developed in this paper can also be applied to find the corrections to gravitational waves sourced by modifications to the Epstein-Wagoner moments~\cite{GravitationalRadiationFromCompactBinarySystems}.
This in turn would allow for tests using both binary pulsar system~\cite{TimingTheBinaryPulsar} and direct gravitational wave detection~\cite{BinaryBlackHoleMergersInTheFirstAdvancedLIGOObservingRun}, complimenting existing work~\cite{PulsarConstraintsOnScreenedModifiedGravity}.
It would furthermore be of interest to use our method to examine the power emitted by the quartic galileon, which does not converge, as shown in ref.~\cite{GalileonRadiationFromBinarySystems}, to test the convergence and consistency of the result with a different expansion prescription.
We leave these analyses for future work.

\acknowledgments

We thank Daniel Thomas, Ignacy Sawicki, Justin Khoury, and Wayne Hu for useful discussions.
This work was supported by the STFC Consolidated Grant for Astronomy and Astrophysics at the University of Edinburgh. L.L.~also acknowledges support from a SNSF Advanced Postdoc.Mobility Fellowship (No.~161058).
Please contact the authors for access to research materials.

\appendix
\section{Parameterised post-Newtonian formalism}
\label{app:PNGR}

The PPN formalism is a parameterisation of the low-energy static regime for generalised gravitational interactions that enables theory-independent tests of gravity in adequate astrophysical regions.
We shall present here a pedagogical primer on the PPN expansion for the unfamiliar reader.
Section~\ref{app:PNGR1} gives a brief review of the low-energy expansion for GR to order $(v/c)^4$. 
This illustrates the generic method by which such an expansion is performed and is easily generalised.
We then present the PPN formalism through comparison to the low-energy expansion of GR to order $(v/c)^4$ in section~\ref{app:PNGR2}.

\subsection{Post-Newtonian expansion of General Relativity}
\label{app:PNGR1}

The low-energy static limit is suitable for describing weak gravitational phenomena which evolve slowly and has been used extensively to test GR in the Solar System~\citep{TheConfrontationBetweenGeneralRelativityAndExperiment}.
For a comprehensive discussion of the expansion and on how to proceed for a range of gravitational theories other than GR, we refer to ref.~\citep{TheoryAndExperimentInGravitationalPhysics}.

We assume that the asymptotic metric far from the system under consideration is Minkowski for some period of time where the cosmological evolution of the metric can be neglected. 
The metric is then expanded about its asymptotic form in orders of $v/c$,
\begin{align}
g_{\mu\nu} = \eta_{\mu\nu} + h_{\mu\nu} = \eta_{\mu\nu} + h_{00}^{(2)} + h_{ij}^{(2)} + h_{0j}^{(3)} + h_{00}^{(4)} \,,
\end{align}
where the indices $i$ and $j$ run from 1 to 3.
The perturbations to the metric are $h_{\mu\nu}^{(i)}$, where the superscript indicates that the field components are of the order $(v/c)^i$.
We will denote the order as $\mathcal{O}_\emph{PN}(i)$ and let $c=1$.
Note that we will keep the gravitational constant $G$ explicit as we make us of it in the main body of the paper.

The virial relation indicates that for the Newtonian potential $U \approx v^2$ and so is of the order $\mathcal{O}_\emph{PN}(2)$.
Further, the Poisson equation indicates that the matter density $\rho$ is of the same order as $U$.
We consider the matter content as a perfect non-viscous fluid.
The pressure in the Solar System is comparable to the total gravitational energy $\rho\,U$, and so we have $p/\rho \approx \mathcal{O}_\emph{PN}(2)$.
Similarly, the specific energy density $\Pi$ is also of the order of the Newtonian potential. 
Hence we can count orders using
\begin{equation}
U \approx v^2 \approx p/\rho \approx \Pi \approx \mathcal{O}_\emph{PN}(2) \,.
\end{equation}
The stress-energy tensor for the fluid to $\mathcal{O}_\emph{PPN}(4)$ is
\begin{align}
\label{eq:stressEnergyTensor00}
T_{00} &= \rho [1 + \Pi + v^2 - h^{(2)}_{00}] \,, \\
\label{eq:stressEnergyTensor0i}
T_{0i} &= - \rho\, v^i \,, \\
\label{eq:stressEnergyTensorij}
T_{ij} &= \rho\, v^i v^j + p\, \delta^{ij} \,.
\end{align}
As we consider a system that evolves slowly in time, we can approximate $d/dt \approx 0$. 
Writing the total derivative in terms of partial derivatives, it is clear that $\partial_t + \overline{v}\cdot\overline{\nabla} \approx 0$.
This relation indicates that time derivatives are of one order higher than the spacial derivatives.

For computational convenience, we adopt the standard post-Newtonian gauge where the $ij$ components of the metric are diagonal and isotropic. 
For GR this gauge choice is specified by 
\begin{align}
\label{eq:GRGuage1}
h^\mu_{ i,\mu} - \frac{1}{2}h^\mu_{\mu,i} &= 0 \,,\\
\label{eq:GRGuage2}
h^\mu_{ 0, \mu} - \frac{1}{2}h^\mu_{\mu,0} &=  -\frac{1}{2}h_{00,0} \,,
\end{align}
where i runs from 1 to 3 and indices are raised and lowered with the Minkowski metric.
In employing these conditions, it is important to remember that the terms compared must be of the same order and that time derivatives increase the order. 
These gauge conditions may change in modified gravity theories to ensure that the metric is diagonal and isotropic.

The Ricci tensor to quadratic order for the $00$ component and linear order for the $ij$ and $0j$ components are 
\begin{align}
\label{eq:R00expansion}
R_{00} =& -\frac{1}{2}\overline{\nabla}^2 h_{00} - \frac{1}{2}(h_{jj,00} - 2h_{j0,j0}) + \frac{1}{2}h_{00,j}(h_{jk,k} - \frac{1}{2}h_{kk,j}) \nonumber \\ 
&- \frac{1}{4}\vert \overline{\nabla} h_{00} \vert ^2 + \frac{1}{2}h_{jk}h_{00,jk} \,,  \\
\label{eq:R0jexpansion}
R_{0j} =& -\frac{1}{2}(\overline{\nabla}^2 h_{0j} - h_{k0,jk}+h_{kk,0j} - h_{kj,0k}) \,, \\
\label{eq:Rijexpansion}
R_{ij} =& -\frac{1}{2}(\overline{\nabla}^2 h_{ij} - h_{00,ij} + h_{kk,ij} - h_{ki,kj} - h_{kj,ki} ) \,.
\end{align}
The $v/c$ expansion is algebraically easiest to perform on the contracted Einstein equations, 
\begin{equation}
R_{\mu\nu} = 8 \pi G ( T_{\mu\nu} - \frac{1}{2}g_{\mu\nu} T) \,.
\end{equation} 

To $\mathcal{O}_\emph{PN}(2)$ we have $R_{00} = -\frac{1}{2}\overline{\nabla}^2 h^{(2)}_{00}$, $T_{00} = -T = \rho$ and $\eta_{00} = -1$. 
The Einstein equations for the 00 component reduce to
\begin{align}
-\frac{1}{2}\overline{\nabla}^2 h^{(2)}_{00} =& 4\pi G \rho \,, \nonumber \\
h^{(2)}_{00} =& 2GU \,,
\end{align}
where we have identified the Newtonian potential $U$.
For $R_{ij}$ we employ the gauge condition~\eqref{eq:GRGuage1} to render the equation diagonal and isometric.
Using that $T_{ij} = 0$ to $\mathcal{O}_\emph{PN}(2)$, the field equation then becomes
\begin{align}
4 \pi G \rho \delta_{ij} =& -\frac{1}{2}(\overline{\nabla}^2 h^{(2)}_{ij} - h^{(2)}_{00,ij} + h^{(2)}_{kk,ij} - h^{(2)}_{ki,ki} - h^{(2)}_{kj,ki} ) \nonumber\\
=& -\frac{1}{2} \overline{\nabla}^2 h^{(2)}_{ij} \,.
\end{align}
Note that the gauge condition is used twice, once on the differentiation with respect to the $i$ and once with respect to the $j$ coordinate. 
Hence
\begin{equation}
h^{(2)}_{ij} = 2 G U \delta_{ij} \,.
\end{equation}

The metric components $h_{0j}$ are $\mathcal{O}_\emph{PN}(3)$ as can be seen from eq.~\eqref{eq:stressEnergyTensor0i}.
The equation of motion for $h^{(3)}_{0j}$ is given by
\begin{align}
8 \pi G \rho v_j =& -\frac{1}{2}(\overline{\nabla}^2 h^{(3)}_{0j} - h^{(3)}_{k0,jk}+h^{(2)}_{kk,0j} - h^{(2)}_{kj,0k}) \nonumber  \\
=& -\frac{1}{2}\overline{\nabla}^2 h^{(3)}_{0j} - \frac{1}{4}h^{(2)}_{00,0j} \nonumber \\
\label{eq:GRPPN0j}
=& -\frac{1}{2}\overline{\nabla}^2 h^{(3)}_{0j} - \frac{1}{2}GU_{,0j} \,,
\end{align}
for which both gauge conditions~\eqref{eq:GRGuage1} and \eqref{eq:GRGuage2} are used.
To evaluate the time derivative to the correct PN order, we make use of the identity
\begin{equation}
\label{eq:timeDerivativeIntergralAprox}
\frac{\partial}{\partial t} \int \rho' f(\overline{x},\overline{x}') d^3 x = \int \rho' \overline{v}' \cdot \overline{\nabla}' f(\overline{x},\overline{x}') d^3x ( 1 + \mathcal{O}_\emph{PN}(2)) \,,
\end{equation}
where $\rho'\equiv\rho(\overline{x}',t)$ and $v_i' = \frac{\partial x'_i}{\partial t}$.
This identity follows from the vanishing of the total derivative.
We thus can solve eq.~\eqref{eq:GRPPN0j} by using the Greens function for the Poisson equation and defining the PPN potentials
\begin{align}
\label{eq:Vi}
 V_i\equiv& \int \frac{\rho' v_i'}{|\overline{x}-\overline{x}'|} d^3x' \,,\\
\label{eq:Wi}
 W_i\equiv& \int \frac{\rho'[\overline{v}'\cdot(\overline{x}-\overline{x}')](x-x')_i}{|\overline{x}-\overline{x}'|^3}d^3x' \,,
\end{align}
to recover the solution
\begin{equation}
h^{(3)}_{0j} = -\frac{7}{2}G V_j - \frac{1}{2}G W_j \,.
\end{equation}

For the  $\mathcal{O}_\emph{PN}(4)$ component of $g_{00}$, we need to use the Ricci tensor up to quadratic order.
Using the gauge conditions, the Ricci tensor becomes 
\begin{align}
\label{eq:R00order4PN}
R_{00}^{(4)} &= -\frac{1}{2}\overline{\nabla}^2 h_{00}^{(4)} - \frac{1}{2}(\overline{\nabla}h_{00}^{(2)})^2 + \frac{1}{2} h_{jk}^{(2)} h^{(2)}_{00,jk} \nonumber \\
&= -\frac{1}{2}\overline{\nabla}^2( h_{00}^{(4)} + 4(\overline{\nabla}^2)^{-1}[(\overline{\nabla}U)^2 - U\overline{\nabla}^2 U]) \nonumber \\
&= -\frac{1}{2}\overline{\nabla}^2( h_{00}^{(4)} + 4(\overline{\nabla}^2)^{-1}[\frac{1}{2}\overline{\nabla}^2U^2 - 2U \overline{\nabla}^2 U]) \nonumber \\
&= -\frac{1}{2}\overline{\nabla}^2(h_{00}^{(4)} + 2U^2 - 8 \Phi_2) \,.
\end{align} 
In the second line we have factored out the Laplacian and inserted the potentials defined earlier.
In the third line we have employed the product rule and in the last line we have used the linearity of the inverse of the Laplacian, defining the PPN potential
\begin{equation}
\label{eq:Phi2Potential}
\Phi_2 = \int \frac{\rho' U'}{|\overline{x} - \overline{x}'|}d^3x' \,.
\end{equation}

Finally, for the matter contribution, we have the 00 component
\begin{equation}
T_{00}-\frac{1}{2}g_{00}T = \rho \left( v^2 - \frac{1}{2}h_{00}^{(2)} + \frac{1}{2}\Pi  + \frac{3}{2}p/\rho \right) \,.
\end{equation}
We define new potentials for the scalar sources through the use of the  Green's function for the Laplacian to find
\begin{equation}
\label{eq:GRh00O4}
h^{(4)}_{00} = -2G^2U^2 +4G\Phi_1 +4G^2\Phi_2 + 2G\Phi_3 + 6G\Phi_4\,.
\end{equation}
We refer to ref.~\citep{TheoryAndExperimentInGravitationalPhysics} for the expressions of the potentials $\Phi_i$. 

\subsection{Parameterising the post-Newtonian expansion}
\label{app:PNGR2}

In the PN expansion of GR described in section~\ref{app:PNGR1}, we have solved the metric perturbations around flat space up to $(v/c)^4$. 
In doing so we have introduced new potentials that solve a range of Poisson-like equations. 
The prefactors to these potentials are theory dependent, as can be seen by performing such an expansion, for instance, for Brans-Dicke theory (see ref.~\cite{ThePostNewtonianEquationsOfHydrodynamicsInTheBransDickeTheory}).

To allow for a theory-independent method of testing gravity, a parameterisation of the metric expansion can be performed. 
This is done in the parameterised post-Newtonian formalism, for which the metric is
\begin{align}
\label{eq:PPNMetric00}
g_{00} = & -1 + 2GU - 2 \beta G^2U^2 - 2 \xi G\Phi_W + (2 \gamma +2 + \alpha_3 + \zeta_1 - 2\xi) G \Phi_1  \nonumber \\
&+ 2(3\gamma - 2 \beta +1 + \zeta_2 + \xi) G^2 \Phi_2 + 2(1+\zeta_2)G\Phi_3 + 2(\gamma +3 \zeta_4 - 2 \xi)G \Phi_4 \nonumber \\
&-(\zeta_1 - 2 \xi)G\mathcal{A} - (\alpha_1 - \alpha_2 - \alpha_3)G w^2 U - \alpha_2 w_i w_j G U_{ij} + (2 \alpha_3 - \alpha_1) w^i G V_i, \\
\label{eq:PPNMetric0i}
g_{0i} = &- \frac{1}{2}(4 \gamma +3 + \alpha_1 - \alpha_2 + \zeta_1 - 2 \xi)GV_i - \frac{1}{2}(1+\alpha_2 - \zeta_1 + 2 \xi)G W_i \nonumber \\
& - \frac{1}{2}(\alpha_1 - 2 \alpha_2)G w_i U - \alpha_2w_jG U_{ij} , \\
\label{eq:PPNMetricij}
g_{ij} = &(1+2\gamma G U) \delta_{ij},
\end{align}
where the PPN parameters are $\gamma$, $\beta$, $\xi$, $\alpha_i$ and $\zeta_i$, and $w^i$ is the velocity of the system with respect to a universal rest frame. 
For a definition of the potentials used in eqs.~(\ref{eq:PPNMetric00}) to (\ref{eq:PPNMetricij}), we refer to ref.~\citep{TheoryAndExperimentInGravitationalPhysics}.
There are several potentials that did not appear in the PPN expansion of GR, such as $\mathcal{A}$. 
These potentials are found, for example, in the expansion of vector-tensor or bimetric theories, \cite{TheoryAndExperimentInGravitationalPhysics}.

From a comparison to the results presented in section~\ref{app:PNGR1}, one finds the GR values $\gamma = \beta = 1$ with all other parameters vanishing.
For a list of the values predicted for these parameters in other gravity theories, see ref.~\cite{TheConfrontationBetweenGeneralRelativityAndExperiment}.
A strength of the PPN formalism is that it directly relates physical effects to each parameter (see, e.g., tab.~\ref{table:1}).
The measurement of those parameters can be compared to predictions for a given theory and used to constrain them, such as with the change in the Shapiro time delay (section~\ref{sec:Shapiro}).
It should be noted, however, that traditionally these parameters are constants, which differs from the results in this work due to the screening effects (see sec.~\ref{sec:CubicPPNSol} and \ref{sec:cham}).

\section{Properties of expansion in the scaling parameter}

In section~\ref{sec:Method}, we have used that the extremal values in a set $Q_j$ obtained from taking an $\alpha$ limit in the scalar field equation are functions of only $q_{i<j}$, where $i$ and $j$ denote the orders of the $\alpha$-correction.
Furthermore, we have used that the higher-order corrections $\psi^{(q_{i>1})}$ obey linear field equations with non-linear equations restrained to the leading order $\psi^{(q_1)}$.
We will discuss these aspects in more detail in appendices~\ref{app:HigherOnLower} and~\ref{app:HigherLinearEquations}, respectively.

\subsection{Perturbative $\alpha$-corrections}
\label{app:HigherOnLower}

In the following, we aim to show that the value of $q_1$ in the expansion of $\phi$ to all orders, given by eq.~\eqref{eq:phiHigherOrder}, is the same as the value $q$ in eq.~\eqref{eq:PhiExpansionLeadingOrder}, where only the term at leading order in $\alpha$ is considered.
That is, $q_1$ is not affected by higher-order $\alpha$-corrections.
For this purpose, let $Z$ denote all fields of a modified gravity theory other than the scalar field $\phi$, for example the metric, and consider a field equation of the form
\begin{equation}
\label{eq:appGenericFieldEquation}
F(Z, \phi, \partial \phi) = T/M_p^2 \,,
\end{equation}
where the source $T$ does not depend on any of the fields. 
As in section \ref{sec:Method}, we will only consider the first derivative of the scalar field in the equations of motion. 
The argument is easily generalised to second or higher derivatives and so for simplicity we will not consider them.

For ease of reference, we repeat the expansion \eqref{eq:phiHigherOrder},
\begin{equation}
\label{eq:appPhiHigherOrder} 
\phi = \phi_0\left(1+\sum_{i=1}^\infty \alpha^{q_i} \psi^{(q_i)}\right) \, .
\end{equation}
As we want the $i^{th}$ term to be larger than the $(i+1)^{th}$, we need an ordering of the exponents $q_i$: for the $\alpha\to \infty$ limit, the ordering is $0\geq q_1 > q_i > q_{i+1}$ and for $\alpha \to 0$, $0\leq q_1 < q_i < q_{i+1}$ for integers $i>1$.
In the following, we shall specify to the limit $\alpha\to \infty$ so that $q_i > q_{i+1}$.
An analogous discussion can, however, be made for the limit $\alpha\rightarrow0$.

We will consider the perturbations $\psi^{(q_i)}$ to be independent fields, so that $F$ is a function of many variables, $\{\alpha^{q_i} \psi^{(q_i)}\, \alpha^{q_i} \partial \psi^{(q_i)}\}$. 
We then Taylor expand $F$ in all $\psi^{(q_i)}$ variables for $i>1$, 
\begin{equation}
\label{eq:FExpansion}
F = F \vert_{\overline{\psi}=0} + \sum_{i>1} \alpha^{q_i}[ (\delta_{\alpha^{q_i}\psi^{(q_i)}} F)\vert_{\overline{\psi}=0} \psi^{(q_i)} + (\delta_{\alpha^{q_i}\partial \psi^{(q_i)}}F)\vert_{\overline{\psi}=0}\partial \psi^{(q_i)}]  + \ldots  \,,
\end{equation}
where the ellipsis contain quadratic and higher order terms, and $\overline{\psi} = (\psi^{(q_2)}, \psi^{(q_3)}, ...)$. 
Importantly, $F$ and its functional derivatives are still only functions of $\alpha^{q_1}\psi^{(q_1)}$ and $Z$ due to the expansion, and $F \vert_{\overline{\psi}=0} = F(X, \psi^{(q_1)}, \partial \psi^{(q_1)})$.

When only considering the first-order correction in the expansion~\eqref{eq:appPhiHigherOrder}, one finds $q_1$ by examining the field equation corresponding to $F \vert_{\overline{\psi}=0}$, which is independent of all $\psi^{(i>2)}$ and hence $q_{i>2}$. 
Thus, we first wish to ensure that this result is replicated when including higher-order corrections.

Let $Q_1$ denote the set of all values that $q_1$ can take such that an exponent of $\alpha$ in $F$ will be equal to zero. 
Then this set will contain elements found from $F \vert_{\overline{\psi}=0}$ which will be the same values found when considering the expansion \eqref{eq:PhiExpansionLeadingOrder}.
However, terms such as $(\delta_{\alpha^{q_i}\psi^{(q_i)}} F)\vert_{\overline{\psi}=0} \psi^{(q_i)}$ give rise to values for $q_1$ that are functions of $q_i$ for $i>1$.
We shall show that the minimum value of the set $Q_1$, and hence the value that $q_1$ takes, is independent of any $q_{i>1}$ and hence $q_1$ coincides with the value found using eq.~\eqref{eq:PhiExpansionLeadingOrder}.

Consider a generic coefficient in the Taylor expansion, $((\prod \delta_{\alpha^{q_i}\psi^{(q_i)}}^{m_i} \delta_{\alpha^{q_i}\partial\psi^{(q_i)}}^{n_i })F)\vert_{\overline{\psi} = \overline{\partial\psi} = 0}$ with finitely many non-vanishing natural numbers $m_i$ and $n_i$.
For such a term not to vanish through performing the functional derivatives on $F$ and the evaluation at $\overline{\psi} = \overline{\partial\psi} = 0$, then in the expansion of $F$, it must be that
\begin{equation}
\label{eq:expansionTerm}
F \supset f(\psi^{(q_1)}, \partial \psi^{(q_1)}) \prod (\alpha^{q_i}\psi^{(q_i)})^{m_i} (\alpha^{q_i}\partial\psi^{(q_i)})^{n_i}
\end{equation}
for some function $f$.
But as we are using the expansion \eqref{eq:appPhiHigherOrder}, there must also be a term 
\begin{equation}
\label{eq:expansionTermOrigin}
F \supset f(\psi^{(q_1)}, \partial \psi^{(q_1)}) \prod (\alpha^{q_1}\psi^{(q_1)})^{m_i} (\alpha^{q_1}\partial\psi^{(q_1)})^{n_i}.
\end{equation}

The exponents of $\alpha$ found from both eq.~\eqref{eq:expansionTerm} and eq.~\eqref{eq:expansionTermOrigin} are then $p + \sum_{i>1} (m_i + n_i) q_i$ and $p + \sum_{i>1} (m_i + n_i) q_1$, respectively, where the exponent $p$ is obtained from $f$. 
The value of $q_1$ can be solved for implicitly by letting $p = p' + q_1$ as well as using the symmetric nature of eq.~\eqref{eq:expansionTerm} and \eqref{eq:expansionTermOrigin}.
Thus
\begin{equation}
Q_1 \supset \left\lbrace - p' - \sum(m_i+n_i)q_1, - p' - \sum(m_i+n_i)q_i \right\rbrace.
\end{equation}
The value that $q_1$ must take is less than or equal to the minimum of these two combinations. 
Using the ordering that $0 \geq q_1 > q_i$ $\forall i>1$, it is clear that the smaller of the two terms is $- p' - \sum(m_i+n_i)q_1$.
Hence for any generic term in the expansion such as in eq.~\eqref{eq:expansionTerm}, the value of $q_1$ which it predicts as a function of all $q_i$ is bounded below by a value that is only a function of $q_1$ from a term such as eq.~\eqref{eq:expansionTermOrigin}.

Thus we have found that when including the higher-order corrections in the expansion~\eqref{eq:phiHigherOrder}, $q_1$ does not become dependent on higher-order exponents $q_{i>1}$.
Moreover, its value remains the same as that obtained from only considering the leading-order term, eq.~\eqref{eq:PhiExpansionLeadingOrder}.
A similar argument can be made for $q_{i>1}$ being independent of $q_{j>i}$, as the key requirement is the ordering of the exponents.

\subsection{Linearity of higher-order $\alpha$-corrections}
\label{app:HigherLinearEquations}

Finally, we aim to show that the field equations for the higher-order perturbations $\psi^{(i>1)}$ are linear for the generic field equation~\eqref{eq:appGenericFieldEquation}.
When finding the field equation for the perturbation $\psi^{(q_1)}$, there are terms within $F \vert_{\overline{\psi}=0}$ that vanish when taking a limit of $\alpha$ as otherwise $\psi^{(q_1)}$ solves the full field equation.
Moreover, only terms that vanish when taking a limit of $\alpha$ remain upon inserting the first two terms of the summation~\eqref{eq:appPhiHigherOrder} into the generic field equation~\eqref{eq:appGenericFieldEquation} and using that $\psi^{q_1}$ solves a field equation which corresponds to the terms proportional to $\alpha^0$.
The remaining terms that are not functions of $\psi^{q_2}$ are those that will source its field equation. 
The slowest vanishing source upon taking a limit is the source that $\psi^{q_2}$ must balance. 

Suppose that this source term takes the form $\alpha^{t}F_t(\psi^{(q_1)}, \partial\psi^{(q_1)})$ for a homogeneous function $F_t$ of order t.
The exponent $q_2$ needs to take a value such that a term in $F$ that is a function of $\psi^{(q_2)}$ is of the same $\alpha$ order as the source term for it to balance the field equation order by order.
In contrast, $q_1$ has to balance the non-vanishing matter source. 
From appendix \ref{app:HigherOnLower} we need only consider the terms in the Taylor expansion $\eqref{eq:FExpansion}$ that are functions of $\psi^{(1,2)}$ or $\partial\psi^{(1,2)}$,
\begin{equation}
\label{eq:FExpansion12}
F = F \vert_{\overline{\psi}=0} + \sum_{n,m}  (\delta^n_{\alpha^{q_2}\psi^{(q_2)}} \delta^m_{\alpha^{q_2}\partial \psi^{(q_2)}} F)\vert_{\overline{\psi}=0} (\alpha^{q_2}\psi^{(q_2)})^n (\alpha^{q_2}\partial \psi^{(q_2)})^m \,.
\end{equation}
Following a similar argument to that in appendix \ref{app:HigherOnLower}, we will show that the value for $q_2$ predicted by a generic term is larger than what a term linear in $\psi^{(2)}$ or $\partial\psi^{(2)}$ would predict for $q_2$.
For a generic coefficient $(\delta^n_{\alpha^{q_2}\psi^{(q_2)}} \delta^m_{\alpha^{q_2}\partial \psi^{(q_2)}} F)\vert_{\overline{\psi}=0}$ to not vanish, there must be a term in $F$ of the form 
\begin{equation}
\label{eq:q2Generic}
F\supset f(\psi_1,\partial\psi_1)(\alpha^{q_2}\psi^{(q_2)})^n (\alpha^{q_2}\partial \psi^{(q_2)})^m \,.
\end{equation}
As we are using the expansion \eqref{eq:appPhiHigherOrder}, the existence of such a term implies that there be a term linear in $\psi^{(2)}$ and $\partial\psi^{(2)}$,
\begin{align}
\label{eq:q2Specific}
F \supset&  f(\psi_1,\partial\psi_1)(\alpha^{q_1}\psi^{(q_1)})^{n-1} (\alpha^{q_1}\partial \psi^{(q_1)})^{m-1} \times \nonumber \\
&\left[  (\alpha^{q_1}\partial \psi^{(q_1)}) (\alpha^{q_2}\psi^{(q_2)}) + (\alpha^{q_1}\psi^{(q_1)})  (\alpha^{q_2}\partial\psi^{(q_2)})  \right] \,.
\end{align}
The exponents found from both eq.~\eqref{eq:q2Generic} and eq.~\eqref{eq:q2Specific} are then $p+(n+m)q_2 $ and $ p+(n+m-1)q_1 + q_2$, respectively, where the exponent $p$ is obtained from $f$. 
Again, let $ p = p' + q_2$. Then we can find the values of $q_2$ that these terms contribute to $Q_2$ as 
\begin{equation}
Q_2 \supset \left\lbrace t-p'-(n+m)q_2, t-p'-(n+m-1)q_1 - q_2 \right\rbrace \,.
\end{equation}
Restricting to the limit $\alpha\to\infty$ and using the ordering $0 \geq q_1 > q_2$, we are left with the value for $q_2$ from eq.~\eqref{eq:q2Generic} being bounded below by the value from eq.~\eqref{eq:q2Specific}.
Hence, any term non-linear in $\psi^{(q_2)}$ or $\partial\psi^{(q_2)}$ provides a value for $q_2$ which is bounded below by a term which is linear in $\psi^{(q_2)}$ or $\partial\psi^{(q_2)}$, and hence any term that contributes to the field equation for $\psi^{(q_2)}$ will be linear in these quantities. The analogous conclusion holds in the $\alpha\to0$ limit.

This argument again can be generalised to the $i^\emph{th}$ case through use of the ordering. 
The result that the field equations for $\psi^{(q_2)}$ are linear may be intuitive as it comes from a Taylor expansion, but we have shown it explicitly here for rigour.

\bibliographystyle{JHEP} 

\bibliography{Scaling_The_Horndeski_Action}

\end{document}